\documentclass[prd,twocolumn,superscriptaddress,showpacs,preprintnumbers,amsmath,amssymb, floatfix1]{revtex4}
\usepackage{amsmath,bm,amsfonts}
\usepackage[dvipsnames]{color}
\usepackage[pdftex]{graphicx}
\usepackage{epstopdf}

\def\msun{$M_{\odot}$}
\usepackage{graphicx}
\usepackage{dcolumn}
\usepackage{bm}
\usepackage{multirow}

\usepackage{slashbox,booktabs}

\newcommand{\s}[1]{{\scriptsize #1}}

\newcommand{\comment}[1]{}

\begin{document}

\title{Towards low-latency real-time detection of gravitational waves from compact binary coalescences in the era of advanced detectors}

\author{Jing Luan}
 \affiliation{
Division of Physics, Mathematics, and Astronomy, Caltech, Pasadena, CA 91125, USA}
\author{Shaun Hooper}
\affiliation{Australian International Gravitational Research Centre,  School of Physics, University of Western Australia, 35 Stirling Hwy, Crawley, WA 6009, Australia}
\affiliation{International Centre for Radio Astronomy Research,  School of Physics, University of Western Australia, 35 Stirling Hwy, Crawley, WA 6009, Australia}
\author{Linqing  Wen}
\email{linqing.wen@uwa.edu.au}
\affiliation{Australian International Gravitational Research Centre,  School of Physics, University of Western Australia, 35 Stirling Hwy, Crawley, WA 6009, Australia}
\affiliation{International Centre for Radio Astronomy Research,  School of Physics, University of Western Australia, 35 Stirling Hwy, Crawley, WA 6009, Australia}
\author{Yanbei Chen}
\email{yanbei@tapir.caltech.edu}
 \affiliation{
Division of Physics, Mathematics, and Astronomy, Caltech, Pasadena, CA 91125, USA}

\begin{abstract}  
Electromagnetic (EM) follow-up observations of gravitational wave (GW) events will {help shed} light on the nature of the sources, and more can be learned if the EM follow-ups can start as soon as the GW event becomes observable.  In this paper, we propose a computationally efficient time-domain algorithm  capable of detecting gravitational waves (GWs)  from coalescing binaries of compact objects with nearly zero time delay.  In case when the signal is strong enough, our algorithm also has the flexibility to trigger EM observation {\it before} the merger.  The key to the efficiency of our algorithm arises from the use of chains of so-called Infinite Impulse Response (IIR) filters, which filter time-series data recursively. Computational cost is further reduced by a template interpolation technique that requires filtering to be done only for a much coarser template bank than otherwise required to sufficiently recover optimal signal-to-noise ratio.  Towards future detectors with sensitivity extending to lower frequencies, our algorithm's computational cost is shown to increase rather insignificantly compared to the conventional time-domain correlation method.  Moreover, at latencies of less than hundreds to thousands of seconds, this method is expected to be computationally more efficient than the straightforward frequency-domain method.
\end{abstract}
\date{\today}
\maketitle
\section{Introduction}

Coalescences of neutron-star (NS) binaries are primary sources for ground-based gravitational-wave detectors.  It has been estimated that Advanced LIGO may be able to detect 10 to 100 such events per year \cite{rate}.  The mergers of neutron star binaries are also possible progenitors of short hard $\gamma$-ray bursts.  
Although these bursts are believed to be mostly beamed away from us, the prompt emission and afterglow they induce in X-ray, optical, infrared and radio frequency bands may well be less beamed, and therefore be visible to us~\cite{fox05, nakar07}.  
If a statistically significant gravitational-wave {\it trigger} can be obtained before or right after such a coalescence,  electromagnetic (especially optical) observatories can then be alerted to search for possible prompt and afterglow emissions  --- such follow-up observations are likely able to resolve whether these mergers are indeed the progenitors of short hard $\gamma$-ray bursts, and provide further knowledge about the nature of these events.

Currently, neutron star - neutron star coalescence signals are being searched for in gravitational-wave data using the  matched filtering technique \cite{finn92, cutler94}, which calculates the correlation of {\it data} with {\it theoretical templates} weighted by noise.  
In order to reduce the computational cost, current search pipelines use a frequency-domain method, which gathers a long stretch of time-series data containing $O(N)$ points (the duration of which should be longer than  the longest possible signal), then uses a Fast-Fourier-Transform (FFT) algorithm to search for all possible signals that end within this stretch of data, with a cost of $O(N \log N)$, as opposed to the $O(N^2)$ required by a one-by-one search over merger time.  
Such a trick, although efficient, implies that we cannot start analyzing the data until the collection finishes.  

Unless significant changes from current frequency-domain analysis method are made, the latency caused by data collection will compromise our ability to obtain a trigger with the shortest possible delay after the merger, and will totally prevent us from obtaining the trigger before the merger.   At least two efforts are underway to suppress latencies for coalescence signals,  the Multi-Band Template Analysis (MBTA)~\cite{MBTA} and the Low-Latency Online Inspiral Detector (LLOID)~\cite{LLOID}.  MBTA is a two-band frequency-domain search method while LLOID provides an infrastructure that accommodates either time or frequency domain searches.   The time-domain aspect of the LLOID pipeline based on Finite-Impulse-Response (FIR) filters~\cite{FIR} is described in a parallel paper \cite{LLOID}.  Note for a different search of short gravitational waves of unknown waveforms,  a program  has been set up to analyze available detector data  in near real-time and  seek for optical counterpart of candidate events \cite{loocup08}. 

In this paper, we propose a straightforward and efficient time-domain search algorithm, which allows zero and even negative latency (i.e., obtaining trigger before the merger if {the signal-to-noise ratio (SNR)} condition and other consistency conditions are met) in the most natural way.  Admittedly, without the savings made available by FFT, the computational cost of a straightforward implementation using FIR filters can be formidable.  {In the correlation calculation}, each template contains a large number of wave cycles, and there exists a large number of templates --- and both these numbers increase dramatically with the lowering of the minimum frequency cutoff $f_{\rm min}$ (Table~\ref{basic_info} ).   {This poses  serious computational challenge for detecting GWs from compact object coalescence for future GW detectors.}  

We propose two techniques that can dramatically increase the computational efficiency for  time-domain searches of GWs from coalescing binaries of compact objects in real-time, and make it feasible for future detectors with frequency cut-offs at as low as $f_{\rm min} =3\, \mbox{Hz}$.  The first technique uses the well-known Infinite Impulse Response, or IIR filters~\cite{FIR}, which can be computed with much higher efficiency than FIR filters.  We propose to filter the data using a bank of IIR filters, the sum of which approximates each individual binary  {coalescence} waveform template.  The second technique reduces the number of templates by an interpolation technique that applies to the proposed IIR filter method.   In this approach, we first divide the bank of IIR filters associated with each template into sub-groups, and then reconstruct the filter outputs of a fine template bank by recombining the filter outputs from each of these sub-groups with appropriate complex coefficients and time delays.  {This is similar to the generic multi-band interpolation scheme used in MBTA and LLOID \cite{MBTA,svd, svd2}.}

\begin{table}[t]
\begin{tabular}{|c|c|c|c|c|}
\hline
Detector & \begin{tabular}{c} $f_{\rm min}$ \\ (Hz) \end{tabular} & \begin{tabular}{c} duration \\ (s) \end{tabular} & $N_{\rm cyc}$ & $\mathcal{N}_{0.98}$ \\
\hline\hline
iLIGO & 40 & 25 & $1.6\times 10^3$ & $1.7\times 10^3$\\ 
\hline
aLIGO & 10 & $1.0\times 10^3$ & $1.6\times 10^4$ & $6.6\times 10^3$ \\
\hline
ET & 3 & $2.5\times 10^4$ & $1.2\times 10^5$ & $2.9\times 10^4$\\
\hline
\end{tabular}
\caption{Basic information for the detection of Newtonian GW signals by  initial, Advanced LIGO and Einstein Telescope.  The columns, from left to right list the names for present and future detectors,  the minimum frequency of the detector, signal duration  and number of wave cycles for a (1.4+14)\msun\ NS-NS binary  [see Sec.~\ref{subsec:nc}], as well as the number of templates required in order to achieve a match of 0.98 for binaries with individual mass of 1 -- 3\,$M_\odot$ [computed from the metric Eq.~\eqref{Ntemplate}]. \label{basic_info}}
\end{table}

Several conventions are used in this paper.   The term {\it latency} refers generally to the delay from the time when a signal arrives at the detector to the time the data containing the signal actually starts to be analyzed.  We specifically focus on the delay starting from the time when the data are ready to be analyzed.   One example of the latency is the delay due to data accumulation before a Fast Fourier Transformation (FFT) can be performed.  The term {\it real time} processing means that data points or data segments are processed (with outputs generated) at a rate that is equal to their input rate.  Floating Point Operation is abbreviated as FLOP (plural FLOPs).   FLOPS and {\it flops} are used interchangeably to stand for Floating Point Operations per Second.  Throughout this paper, we follow the convention of counting each real addition and real multiplication equally as one FLOP.  

This paper is structured as follows.  In Section \ref{basics} , we briefly review the basics of  matched filtering technique and introduce time-domain IIR filters.  In Section \ref{chirpsintoiir}, we use Newtonian-order templates as an example to construct IIR filters, characterize the error involved and calculate the computational cost for each individual template.  In Section \ref{filtersubgroups}, we present an interpolation technique that allows us to use a significantly decreased number of templates for which filter chains must be implemented. In Section \ref{sec:tf}, we make a simple comparison between the computational cost of IIR filtering and the straightforward frequency-domain algorithm.  In Section \ref{conclusion}, we summarize our main conclusions.

\section{Matched Filtering Technique\label{basics}}


The optimal technique to extract a  signal from noisy data when we have reliable theoretical predictions for the signal {\it waveform} is to use {\it matched filtering} \cite{finn92, cutler94}. The output of the matched filtering technique is basically the correlation of data with expected waveforms weighted by noise.  This can be realized in the frequency or time domain. We will give a brief overview of the matched filtering technique, and introduce its frequency-domain implementation and its time-domain approach using the FIR and IIR filters.  

\subsection{Frequency-domain implementation}

\subsubsection{Single Template}
Suppose the output of the interferometer $h$ is a sum of noise $n$ and, if exists, a signal $s$:
\begin{equation}
  \label{eqh}
  h = n + s
\end{equation}
For the moment, let us assume that $s$ is a single known {\it waveform}.  In Eq.~\eqref{eqh}, we have intentionally left out the arguments of the functions $h$, $n$, and $s$, which reflects the point of view that each of them can be equivalently {\it represented} both in the time and frequency domain.  More specifically, we use the following convention for Fourier transform, which relates $h(t)$ and $\tilde h(f)$ (we shall use tilde to emphasize a frequency-domain representation):
\begin{equation}
  \tilde h(f)\equiv \int_{-\infty}^{\infty} dt e^{i 2\pi f t} h(t).
\end{equation}
The power spectral density of $n(t)$ is denoted by $S_h (f)$, which is defined by
\begin{equation}
  E[\tilde n(f) \tilde n^*(f')]=\frac 1 2 \delta (f-f') S_h (f).
\end{equation}
Here we use one-sided spectral density, $E[\ ]$ denotes the expectation value over an ensemble of realizations of the noise and ``$*$" denotes complex conjugation. $S_h(f)=S_h(|f|)$ as the noise in the time domain $n(t)$ is real.

In order to extract $s$ from $h$, we perform  {\it filtering}, which consists of taking the {\it inner product} between data $h$ and {\it template} $u$, forming a {\it filter output} of $y$:
\begin{equation}
  y = \langle h | u\rangle = \langle s | u\rangle + \langle n| u\rangle
  \label{MF_f}
\end{equation}
Here we define inner product as
\begin{eqnarray}\label{innerproduct}
  \langle a|b\rangle&\equiv& 2 \int_{0}^{\infty} df \frac{\tilde a^*(f)\tilde b(f)+\tilde a(f)\tilde b^*(f)}{S_h(f)}\nonumber\\
  &=&4 \mathrm{Re} \left[ \int_0^{\infty}\frac{\tilde a^* (f) \tilde b(f)}{S_h(f)}\right].
\end{eqnarray}
In $y$, we have a signal component $\langle s | u\rangle$ and a noise component $\langle n | u\rangle$ which fluctuates around zero.  If $s$ has a substantially high amplitude and if the template $u$ is appropriate, the signal component $\langle s | u\rangle$ in $y$ will raise to a high value that merely random fluctuation of $\langle n | u\rangle$ is very unlikely to account for.  As a consequence, we can impose a threshold on $y$ --- an incidence with $y$ higher than the threshold is viewed as a detection of a signal.  { The detection efficiency depends on the signal-to-noise ratio (SNR) defined generally as
\begin{equation}
\rho= \frac{y(n=0) - E[y(s=0)]}{\sigma_{ y(s=0)}}, 
\end{equation}
where $\sigma_{ y(s=0)}$ is the standard deviation of the filter output when data contain noise only.    Assuming zero-mean  Gaussian noise,  we have for Eq.~\eqref{MF_f} } 
\begin{equation}\label{SNR}
  \rho\equiv \frac{\langle s|u\rangle} {\sqrt{E[| \langle n|u\rangle|^2]}},
\end{equation}
Note that the SNR does not depend on the normalization of the template $u$, and it is conventional to require that  $\langle u|u\rangle=1$.  In this case,  the cross-correlation of a template with pure noise $\langle n|u\rangle$ is a random variable with zero mean and unity variance. It is easy to show that $E[\langle n|a\rangle\langle n|b\rangle]=\langle a|b\rangle$. So we have
\begin{equation}
  \rho=\langle s|u\rangle.
\end{equation}
According to the Cauchy-Schwarz inequality,
\begin{equation}
  \rho=\frac{\langle s|u\rangle}{\sqrt{\langle s | s \rangle}} \sqrt{\langle s | s \rangle}  \le  \sqrt{\langle s | s \rangle},
\end{equation}
where  equal sign takes place when $u  = \lambda s$ where $\lambda$ is a constant, and normalization of $u$ gives $\lambda = 1/\sqrt{\langle s | s\rangle}$.   This means the {\it optimal} SNR is given by the modulus of the signal, $\langle s | s \rangle$, and the {\it reduction} of SNR due to {imperfectness} of template is given by the {\it match}, which is also equal to unity minus mismatch, $\varepsilon$: 
\begin{equation}
 \frac{\langle s|u\rangle}{\sqrt{\langle u | u\rangle\langle s|s \rangle}}  \equiv 1-\varepsilon.
\end{equation}

\subsubsection{Intrinsic and Extrinsic Parameters}
\label{subsubsec:intext}

In reality, templates are not necessarily placed along each parameter dimension.   The maximization of SNR  over certain parameters can be conducted analytically and therefore no templates are needed.  These parameters are called {\it extrinsic parameters}, while those that still have to be searched over one by one are called {\it intrinsic parameters}.

As an example, for any generic waveform $u(t) = Au_0(t-t_c) e^{i\phi_c}$, where $A$ is a real number, $\phi_c$ is the phase difference between $u(t_c)$ and $u_0(t_c)$, and $t_c$ is its {\it ending time}. The ending time $t_c$ is an extrinsic parameter, because as a series of templates $u_0(t-t_c)$ with a variety of $t_c$ are applied to the data $h$,  the SNR
\begin{align}
  &
\rho(t_c) = 4\,\mathrm{Re} \int_{0}^{\infty} \frac{\tilde{h}^*(f)\tilde{u}_0(f)}{S_h(|f|)}e^{i2\pi ft_c} df
  \label{matchedFilter}
\end{align}
can be computed  for all $t_c$ via a Fast Fourier Transform, which cost $O(N\log N)$ FLOPs in the discretized case  where $N$ is the number of data points in the time domain.  This is much faster than computing the correlation for all possible ending times, one by one, which cost $O(N^2)$ operation counts --- and in this way ending time $t_c$ is converted into an extrinsic parameters. The method of Fourier transformation will be discussed in detail in subsection \ref{Fourier Transform}. This process dominates the computational cost for the matched filtering method.  Further analytical optimization are known for the search of the constant phase $\phi_c$.  We assume the process is similar for all methods discussed in this paper and that its computational cost is negligible.

\subsection{Time-domain Approach: FIR and IIR method}
\label{subsecIIR}

For the time-domain filtering we need to obtain a time series of  SNRs as a function of presumed signal arrival time $t$
\begin{eqnarray}
  \label{eqFIR}
  \rho(t)&=& 2\int_{-\infty}^{\infty} \frac{\tilde{h}^* (f)\tilde{u}(f)}{S_h(|f|)} e^{+i2\pi ft}  df\nonumber\\
  &=&\int_{-\infty}^{+\infty} w(t') u(t'-t)dt'
\end{eqnarray}
with
\begin{equation}\label{overwhite}
  w(t)\equiv 2\int_{-\infty}^{\infty}df \frac{\tilde h(f)}{S_h(f)} e^{-i 2\pi f t},
\end{equation}
which can be thought of as ``over-whitened data''; it is a real-valued function of time.  Note that in order to generate the over-whitened data, we need to convolve $h(t)$ with the Inverse Fourier Transform of $1/S_h(f)$, which is a time-symmetric, oscillatory function that decays towards zero when $t$ is much larger than the inverse of the interferometer's bandwidth ($\stackrel{>}{_\sim}100$\,Hz), which is about $\stackrel{<}{_\sim}10\mbox{ms}$. This means the over-whitening process has an inherent latency not much larger than $10\mbox{ms}$, which is negligible compared  to the duration of the signal.

We now discretize the filtering algorithm. The discrete form of Eq~\eqref{eqFIR} becomes,
\begin{equation}
  \rho_k=\sum_{j=-\infty}^k w_j u_{j-k}\Delta t\label{IIR},
\end{equation}
Here we assume $t_k = k \Delta t$, and that $u$ only have support within $t \le 0$. While in principle the waveform $u_k$ could have an infinite support in time, $-\infty <k \Delta t <0$. However, the waveform $u(t)$ is always assumed to begin only after its amplitude reaches sensitivity within the LIGO band. Hence we instead define the waveform to exist on the domain $-N\Delta t \le  t \le 0$, and Eq~\eqref{IIR} becomes,
\begin{equation}
  \rho_k=\sum_{j=k-N}^k w_j u_{j-k}\Delta t
\label{FIR},
\end{equation}
This summation of the product of data and template at each step turns out to be the general form of Finite Impulse Response (FIR) filters. The term {\it finite} comes from the fact that the output $\rho_k$ of the filter (its response) will become exactly zero after $N$ time steps have passed since a single initial impulse in the data. For example, if we assume $w_0=1$, $w_k = 0 $ for $k \neq 0$, then $\rho_k$ will vanish for $k > N$.  As seen from Eq.~\eqref{FIR}, each $\rho_k$ costs $N$ multiplications and $N$ additions to calculate.   This translates into a computational cost, in terms of FLOPs per unit time, of $\sim N/\Delta t$. 

For certain types of waveforms, Infinite Impulse Response (IIR) filters can be used to dramatically reduce computational cost.   The simplest IIR filter is a first-order recursive algorithm,  in which the $k$-th output $y_k$ is a linear combination of the $(k-1)$-th output, $y_{k-1}$ and the $k$-th data, $w_k$:
\begin{equation}
  \label{iirfilter}
  y_k=e^{-(\gamma-i\Omega)\Delta t} y_{k-1}+   w_k \Delta t,
\end{equation}
where $\gamma, \Omega$ are real-valued constants with $\gamma > 0$ to ensure stable solutions.  It can be shown that, e.g., by using tools of Z-transform~\cite{DP75}, as long as $w_k$ does not diverge towards $k\rightarrow -\infty$, then even if the recursion starts at a finite time step, after an initial transient of several times $1/\gamma$, the output of the filter achieves a steady state of  
\begin{equation}\label{IIRdis}
  y_k=\sum_{j=-\infty}^{k} w_j  e^{(\gamma-i\Omega)(j-k)\Delta t} \Delta t.
\end{equation}
Note this is the discretized version of the continuous integration
\begin{equation}\label{IIRcon}
  y(t)=\int_{-\infty}^{\infty} w(t')  e^{(\gamma-i\Omega)(t'-t)} \Theta(t-t') dt'.
\end{equation}
Note first that  Eq.~\eqref{IIRdis} indeed gives an infinite impulse response, because for a data series containing only one impulse, $w_0=1$ and $w_k=0$ ($k\neq0$), the output of the filter, even at very late time steps, never vanishes.  More over, by comparing this with Eq.~\eqref{eqFIR}, the IIR filter can be viewed as a template of a damped sinusoid: 
\begin{equation}
  \label{IIRform}
  u(t)=  e^{(\gamma+i\Omega)t} \Theta(-t),
\end{equation}
where $\Theta(t)$ is the Heaviside function
\begin{eqnarray}
\Theta(t) = \left\{
\begin{array}{ll}
 0, & t\le 0; \\
 \\
 1, & t\ge 0.
 \end{array}
 \right. 
\end{eqnarray}
The IIR filter described above requires only one complex multiplication and one summation per sampling time, which means the computational cost is $ \sim 1/\Delta t$.  

For a simple proof of concept on the computational efficiency of the IIR over the FIR filtering technique, we examine the case when we {\it do} need to filter for a damped sinusoid signal with frequency $\Omega$ and decay rate $\gamma$.  {The data to be filtered has a duration of at least on the order of } $1/\gamma$.  The Nyquist sampling theorem limits the sampling interval to be at most $\sim 1/\Omega$, meaning that the FIR template would have the number of data points several times larger than $\Omega/\gamma$.  Subsequently the computational cost of the FIR in FLOPs per unit time is larger than $\Omega^2/\gamma$.  An IIR filter, on the other hand, only has a cost of $\Omega$, which means the cost of IIR filter is $\gamma/\Omega \sim 1/Q$ times that of FIR filter, where $Q \equiv \Omega/\gamma$ is the quality factor of the damped sinusoid.   As a consequence, if we can convert our waveforms into a sum of a series of of high-$Q$ damped sinusoids, IIR filters can be used over the FIR to dramatically reduce the computational cost.

\section{Construction of IIR filters for an individual inspiral waveform\label{chirpsintoiir}}
\label{strategy}
The simple IIR filter discussed in the previous section 
has the special waveform of a decaying sinusoid [Cf.~Eq.~\eqref{IIRform}].  In this section, we will show that a {\it chain} of IIR filters can be used to ``piece together'' the waveforms of compact binary coalescence.   This is possible because these waveforms are basically sinusoids with slowly varying amplitude and frequency. For simplicity, in this paper, we will restrict ourselves to Newtonian Chirps.

\subsection{The Newtonian Chirp Waveform}
\label{subsec:nc}

The Newtonian-chirp is the leading-order waveform from a compact coalescing binary.  In the time domain it can be written as the real part of the complex expression (see, e.g., \cite{cutler94}, Sec. C),
\begin{equation}\label{newtonian time}
  u(t) \propto (t_c-t)^{-1/4}e^{-i 2(5M_c)^{-5/8}(t_c-t)^{5/8}+i\phi_c} \equiv \mathcal{A}(t) e^{i\Phi(t)}
\end{equation}
where we follow the convention of the Planck unit that sets gravitational constant $G=1$ and the speed of light,  $c=1$,  $M_c$ is the {\it chirp mass} of the binary, 
\begin{equation}
M_c=M\eta^{3/5}
\label{chirp_mass}
\end{equation}
which depends on the   total mass of the binary $M$ and $\eta \equiv m_1 m_2/M^2$, the symmetric mass ratio.  The signal finishes at the ending time $t_c$, and $\phi_c$ is the constant phase at the end time.
Here we have ignored time-independent factors of proportionality in the amplitude, which do not affect template construction.  

We have assigned real-valued functions $\mathcal{A}(t)$ and $\Phi(t)$ to denote the amplitude and phase of the waveform.  Although the actual waveform is the real part of $u(t)$, we have intentionally kept its complex form, because the imaginary part of $u(t)$ represents the waveform of a binary with a phase shift of $\pi/2$ from the real part--- therefore the real and imaginary parts together form a {\it basis} for the {\it linear space of signals of all phases}.  This is a feature of all {\it adiabatic waveforms}, which satisfy 
\begin{equation}
  \label{adiabatic}
  \dot{ \mathcal{A}}/(\Omega  \mathcal{A})\ll1,\quad \dot\Omega/\Omega^2 \ll 1.  
\end{equation}
In other words,  the  amplitude $\mathcal{A}(t)$ and angular frequency $\dot\Phi(t)$ both evolve at rates much slower than the instantaneous frequency $\dot\Phi$.  This  allows us to use the Stationary Phase Approximation (SPA) to compute the Fourier Transform of the waveform in Eq.~\eqref{newtonian time},
\begin{equation}
  \tilde u(f) \propto f^{-7/6} e^{i(A f^{-5/3}+2\pi f t_c+\phi_c-{\pi}/{4})},\quad f>0\, , 
\end{equation}
where
\begin{equation}
  A=\frac{3}{4}(8\pi M_c)^{-5/3} 
\end{equation}
is the intrinsic parameter we need to search for in the case of Newtonian chirp.  Note that when we Fourier-transform the complex signal of Eq.~\eqref{newtonian time}, there is only positive-frequency component, with $\tilde u(f)=0$ for $f<0$. On the other hand, if we took the real part of the signal, we would have $\tilde u(-f) = \tilde u^*(f)$ for $f>0$.


The duration of a coalescence GW signal can be well approximated as a function of chirp mass $M_c$ (Eq.~\eqref{chirp_mass}) and the detector's minimum cut-off frequency $f_{\rm min}$,
\begin{equation} T (M_c,
  f_\mathrm{min})=\frac{647013}{(f_{\mathrm{min}}/\mathrm{Hz})^{8/3}
    (M_c/M_{\odot})^{5/3}} \ \ \mathrm{s}. \label{Tsignal} \end{equation} 
One can see that for a fixed $f_{\rm min}$,  the longest signal duration corresponds to the smallest chirp mass.  The sample signal durations for the initial, advanced and future GW detectors of various $f_{\rm min}$  can be found in Table~\ref{basic_info}, column 3.  It is shown that GWs from a canonical (1.4+1.4)\msun\  NS-NS binary system will have a duration 40 times longer for advanced detectors, and possibly 10000 time longer for the future ET detector than that of the initial detector. 

\subsection{An IIR filter chain}
\label{subsec:chain}

The adiabatic condition in Eq.~\eqref{adiabatic} also implies that the waveform can be divided into {\it constant-frequency intervals}: within each interval it can be approximated as a sinusoid with constant frequency, while neighboring intervals have slightly different frequencies.  This further indicates that we can attempt to write the entire waveform into the sum of a series of damped sinusoids: the frequency of each sinusoid corresponds to a constant-frequency interval, the ending time of the sinusoid corresponds to the ending time of this constant-frequency interval, while the decay time should be comparable to the length of the constant-frequency interval. The amplitude of the decaying sinusoid can be set to be comparable to the amplitude of the original waveform during the corresponding constant-frequency interval. 

 Mathematically, our target is therefore to approximate the signal template $u(t)$ with the sum of a chain of IIR filters [Cf.~Eq.~\eqref{IIRform}], which we  denote by $U(t)$:
\begin{equation}
  U(t) \equiv \sum_{l=1}^{M} B_l e^{(\gamma_l-i\Omega_l)(t-t_l)}\Theta(t_l-t). 
\label{U_t}
\end{equation}
Here the chain consists of $M$ filters; for filter $l$ ($1 \le l \le M$), $B_l$ is the amplitude of the filter $l$, $\Omega_l$ and $\gamma_l$ are the angular frequency and decay rate, and $t_l$ is its ending time.

As a first step, let us determine the relevant portion of the signal that we need to approximate: this is bounded by the low frequency cut-off $f_{\mathrm{min}}$, below which the chirp only contributes negligible signal-to-noise ratio, as well as the high frequency cut-off $f_{\mathrm{max}}$.  The minimum frequency $f_{\mathrm{min}}$ is normally determined by the seismic wall of the detector which is set to be 40\,Hz for initial LIGO, 10\,Hz for Advanced LIGO, and might extend to lower frequencies in future detectors, such as the Einstein Telescope (ET).  
  The maximum frequency $f_{\mathrm{max}}$  is either determined by the end of the Newtonian chirp or the upper end of the detection band.    In this paper we set $f_{\mathrm{max}} = 2000\,{\rm Hz}$.   

Now suppose our Newtonian chirp has a particular value for the intrinsic parameter $A$, and $t_c=0$, $\phi_c=0$.  Let us define  $t_0\equiv t_{\mathrm{ini}}$ as the time at which the instantaneous frequency of the waveform is equal to $f_{\rm min}$ (which means $|t_0| = -t_0$ is the duration of the Newtonian chirp from $f_{\rm ini}$ to coalescence), and incrementally define
\begin{equation}
  \label{Tl}
  t_{l} = t_{l-1}+ T_l\,,\quad \left| \frac 1 2 \ddot\Phi(t_l) T_l^2\right|=\epsilon\ll 1\,,\quad l=1,2,\ldots
\end{equation}
until we reach $t_{M}$, which corresponds to a frequency at or beyond $f_{\rm max}$.  These intervals,
\begin{equation}
[t_0,t_1],\;[t_1,t_2],\;\ldots,\;[t_{M-1},t_{M}]
\end{equation}
will be the constant-frequency intervals described previously. The parameter $\epsilon$ should be substantially less than unity, so that the phase error caused by assuming a constant frequency is significantly less than one radian. 

For $t \in [t_{l-1},t_l]$, we expand $\Phi(t)$ at $t_*=t_l-\alpha T_l$ (where $\alpha$ is an ad hoc parameter to be adjusted later)
\begin{equation}
  \Phi(t)\simeq \Phi(t_l^*)+\dot\Phi(t_l^*)(t-t_l^*)+\frac 1 2 \ddot\Phi(t_l^*)(t-t_l)^2 
\end{equation}
such that the first term is a constant phase, the second term gives a single angular frequency of $\dot\Phi(t_l^*)$, while the third term gives the error of a single-frequency approximation, which will be small if $\epsilon$ is small enough in Eq.~\eqref{Tl}.  
We will then use $\Omega_l \equiv -\dot\Phi(t_l^*)$ as the oscillation frequency of the IIR filter assigned for this constant-frequency interval, and prescribe a complex amplitude of 
\begin{equation}
  B_l\equiv\mathcal{A}(t_l^*)e^{i\Phi(t_l^*) -i\Omega_l(t_l-t_l^*)}\,. \end{equation}
These will assemble into
\begin{eqnarray}
  \label{tempIIR}
  B_l e^{-i\Omega_l (t-t_l)} & =&  \mathcal{A}(t_l^*)e^{i\Phi(t_l^*) +i\dot\Phi(t_l^*)(t-t_l^*)}\nonumber \\ &\approx& \mathcal{A}(t) e^{i\Phi(t)},\quad t_{l-1} \le t \le t_l\,.
\end{eqnarray}

We must still add a Heaviside function and a damping component to modify \eqref{tempIIR} into a form realizable by an IIR filter.  Since the validity of \eqref{tempIIR} is between $t_{l-1}$ and $t_l$, it is natural to have the Heaviside function cut off values for $t > t_l$, and to have the damping component have a time constant comparable to $T_l$, which gradually cuts off the filter at $t \stackrel{<}{_\sim} t_{l-1}$. Prescribing 
\begin{equation}
  \gamma_l = \zeta/T_l\,,
\end{equation}
with $\zeta$ yet another ad hoc parameter, we write
\begin{eqnarray}
  &&U_l(t;A,t_c=0,\phi_c=0)  \nonumber\\
  &\equiv&  B_l e^{-i\Omega_l (t-t_l)-\gamma_l (t_l-t)} \Theta(t_l -t)
\end{eqnarray}
which is our IIR filter for interval $l$, for chirps with parameters $A$, $t_c=0$, $\phi=0$. Summing over all $U_l$, we obtain an IIR chain that approximates the entire {\it complex} chirp signal:
\begin{equation}\label{Ubeforeadjust}
  U(t;A, t_c=0,\phi_c=0)= \sum_{l=1}^M U_l(t;A, t_c=0,\phi_c=0).
\end{equation}
If the sum of the complex filter chain  $U(t)$ indeed approximates the complex chirp signal $u(t)$ [Cf.~Eq.~\eqref{newtonian time}], then the  real and imaginary parts of the output from the filter chain will be good approximations for filtering chirps with $\phi_c=0$ and $\pi/2$, respectively.  

For non-zero $t_c$, we will have to apply
\begin{equation}\label{Utcbeforeadjust}
  U(t;A,t_c,\phi_c=0) \equiv U(t-t_c;A,t_c=0,\phi_c=0)
\end{equation}
Note that having Heaviside Function $\Theta(t-t_c-t_{l})$ within $U_l$  means we have to collect the IIR filter result of  filter $l$ at $t_{l}+t_c$.  The fact that all $t_l$ are negative means all results are obtained {\it before} the coalescence (which happens at $t_c$) and hence IIR filtering itself causes no latency --- except for the small latency due to over-whitening, as stated previously (sec.~\ref{subsecIIR})

\subsection{Filtering for general signal phases and goodness of match}
\label{sec:match}

Since the construction of the IIR filter chain is of an ad hoc nature, we must test how well the resulting IIR filter chain $U$ can approximate the original signal $u$.  A natural candidate would be imposing that the match between the signal $u$ and the template $U$
 \begin{equation}
\rho_{\rm cplx}=   \frac{|\langle u | U\rangle |}{
    \sqrt{\langle u | u\rangle  \langle U | U\rangle  }} 
\end{equation}
must be close to unity. 

However, this needs to be connected to the signal-to-noise ratio achievable by IIR filtering. For doing so, we must first elaborate how to use the output of the complex IIR filtering to recover signals with arbitrary phases.  If we write 
\begin{equation}
u \equiv u_r + i u_i
\end{equation}
with $u_{r,i}$ represent the real and imaginary parts of $u$ in the time domain, and similarly, 
\begin{equation} 
U \equiv U_r + i U_i\, ,
\end{equation} 
then the true signal of arbitrary phase is a linear combination of $u_r$ and $u_i$ written as $A_1 u_r + A_2 u_i$, and we should use a linear combination of the real and imaginary parts of $U$, namely $B_1U_r+B_2U_i$ as the search template.   For any particular coefficients $A_{1,2}$, the optimal overlap is given by 
\begin{equation}\label{rhoIIR}
  \rho_{\mathrm{IIR}}(A_1,A_2)=\max\limits_{B_{1,2}} \frac{ \langle A_1u_r+A_2u_i|B_1U_r+B_2U_i\rangle}{\sqrt{\langle B_1U_r+B_2U_i|B_1U_r+B_2U_i\rangle}}
\end{equation} 
The worst-case scenario is given by a minimization over $(A_1,A_2)$:
\begin{equation}\label{rhoIIRworst}
  \rho_{\mathrm{IIR}}^{\mathrm{worst}}=\min\limits_{A_1,A_2}\frac{\rho_{\mathrm{IIR}}(A_1,A_2)}{\sqrt{\langle A_1u_r+A_2u_i|A_1u_r+A_2u_i\rangle}}.
\end{equation} 
In fact, when the signal and the template are both highly adiabatic, it can be shown that $\rho_{\rm IIR}(A_1,A_2)$ is approximately independent of $A_{1,2}$, and that to a very good accuracy:
\begin{equation}
\rho_{\rm cplx} \approx \rho_{\rm IIR}^{\rm worst}.
\end{equation}
Eq.~\eqref{rhoIIRworst} is therefore used to calculate the goodness of the match of the IIR filter chain.


\begin{table*}[t]
  \begin{center}
    \begin{tabular}{|c||c|c|c|c|c|c|c|c|c|c|c|c|c|c|}
      \hline
\multirow{2}{*}{} & \multirow{2}{*}{\backslashbox{{\tiny Type}}{\tiny Rate}} & 
$S_k$ ($\mathrm{s}^{-1}$) & 16 & 32 & 64 &  128 & $256$ & $512$ & $1024$ & $2048$ & $4096$ &$8192$ 
& \multirow{2}{*}{$N_{\rm tot}$} & Total \\
    \cline{3-13}
& & 
$f/\mbox{Hz}$ & \s{2 -- 4} & \s{4 -- 8} & \s{8 -- 16}  & \s{16 --  32} & \s{32 -- 64} & \s{64 -- 128} & \s{128 -- 256} & {\footnotesize 256 -- 512} & {\footnotesize 512 -- 1024} & {\footnotesize $>$1024} 
& & Cost\\
      \hline\hline
      \multirow{4}{*}{iLIGO} & 
      \multirow{2}{*}{FIR} &
      $N_{\tiny\mbox{FIR,k}}$ 
      & & & & & $4547$ & $3062$ & $965$ & $304$ & $96$ & $30$ & \multirow{2}{*}{9004} & \multirow{2}{*}{20}\\
      \cline{3-13}
 &      & $\mathcal{C}_{{\rm FIR},k} $  & & & &  & 4.7 & 6.3 & 4.0 &2.5 & 1.6  & 1.0 & & \\ \cline{2-15}
& \multirow{2}{*}{IIR}     & $N_{\tiny\mbox{IIR,k}}$ 
& & & & & $71$ & $62$ & $ 34$ & $19$ & $10$ & $4$ & \multirow{2}{*}{200} & \multirow{2}{*}{2.4}\\
     \cline{3-13} 
          &  & $\mathcal{C}_{\rm IIR}$ & & & & & 0.22 & 0.38 & 0.42 & 0.47 & 0.49 &0.39   & &  \\     \hline\hline
      \multirow{4}{*}{aLIGO} & 
      \multirow{2}{*}{FIR} &
      $N_{\tiny\mbox{FIR,k}}$ 
      & & & \s{45835} & \s{30868} & 9723 & 3062 & 965 & 304 & 96 & 30 & \multirow{2}{*}{\s{90883}} &  \multirow{2}{*}{53}\\
      \cline{3-13}
 &      & $\mathcal{C}_{{\rm FIR},k} $  
 & & & 11.7 & 15.8 & 10.0  & 6.3 & 4.0 & 2.5 & 1.6  & 1.0 & &  \\ \cline{2-15}
& \multirow{2}{*}{IIR}     & $N_{\tiny\mbox{IIR,k}}$ 
& & & 220 &  198  & 111 & 62 &34  & 19  & 10 & 4& \multirow{2}{*}{658} & \multirow{2}{*}{3.0}\\
     \cline{3-13} 
          &  & $\mathcal{C}_{\rm IIR}$ 
          & &  & 0.17 & 0.30  & 0.34 & 0.38 & 0.42 &0.47 &0.49   & 0.39 & &  \\     \hline\hline
      \multirow{4}{*}{ET$_{\rm B}$} & 
      \multirow{2}{*}{FIR} &
      $N_{\tiny\mbox{FIR,k}}$ 
      & \s{213010} & \s{311130} & \s{98000} & \s{30868} & 9723  &3062  & 965 & 304 &  96 &40  &  \multirow{2}{*}{\s{667198}} & \multirow{2}{*}{120}\\
      \cline{3-13}
 &      & $\mathcal{C}_{{\rm FIR},k} $  &  13.6 & 39.8  & 25.1 & 15.8 & 10.0 & 6.3 & 4.0 & 2.5 & 1.6  &1.0 &  & \\ \cline{2-15}
& \multirow{2}{*}{IIR}     & $N_{\tiny\mbox{IIR,k}}$ & 392 & 631 & 353 & 198 &  111 & 62 & 34 & 19  &11  & 3 & \multirow{2}{*}{1814} &  \multirow{2}{*}{3.3}\\
     \cline{3-13} 
          &  & $\mathcal{C}_{\rm IIR}$ &  0.08 & 0.24 & 0.27 & 0.30 & 0.34  &  0.38 &  0.42 & 0.47 &  0.54   & 0.29 &  & \\     \hline\hline

    \end{tabular}
    \caption{
    Break-down of number of filters and computational cost (over successive two-fold down-sampling channels) of multi-rate FIR and IIR filtering, of a single template for a $(1.4+1.4)\,M_\odot$ binary for initial, Advanced LIGO and the Einstein Telescope.  See text in Sec.~\ref{subsec:implementation}.  Here computational costs for each type of filtering and for different sampling channels are calculated using Eqs.~\eqref{CCiir}--\eqref{CCfir}, with numerical values quoted in units of MFLOPS or $10^6\,$FLOPS. The minimum overlap is 0.99.
    \label{tab_downsample}
}
  \end{center}
\end{table*}

\subsection{Implementation for $(1.4+1.4) M_\odot$ binaries and initial LIGO}
\label{subsec:implementation}

We first apply the prescription described in Sec.~\ref{subsec:chain} to construct an IIR filter chain for $(1.4+1.4)M_\odot$ binaries for initial LIGO and use  Eq.~\eqref{rhoIIRworst} to test their overlap with the true signals.  We choose (by hand) $\alpha=2.3$, $\epsilon=0.269$ and $\zeta=4$, an overlap of $0.99$ is achieved with  $N_{\mbox{\scriptsize{IIR}}}=200$ IIR filters. 

We next estimate the computational cost required by such IIR filtering.  {We focus on the floating point operation count per unit time required to generate complex outputs from the sum of individual IIR filter outputs of Eq.~\eqref{iirfilter}.}  Here we assume the maximum sample rate for compact-binary coalescence data analysis is 8192\,Hz, with 2$\times$ down-sampling applied successively to provide channels with sample rates of 4096\,Hz, 2048\,Hz, \ldots, 256\,Hz. 
The IIR filter bank is divided into $6$ groups, each corresponding to a frequency band of $2^{k+5}$--$2^{k+6}\,$Hz, for $k=0,1,...,5$.  For filters in group $k$, we assume they are applied to the channel with sample rate of 
\begin{equation}
  S_k = 2^{k+8}\,\mathrm{Hz}\,.
\end{equation}
In Table~\ref{tab_downsample}, we list the actual number of IIR filters required {to achieve a minimum overlap of 0.99} at different frequency band with downsampling technique.  For comparison,  we list the corresponding numbers for the FIR method also applied with downsampling technique. 

At each time step, each IIR filter needs to perform a total of 12 real-number multiplications and additions  namely: 4 real-number multiplications plus 2 real-number additions for multiplying the current output by the complex recursive coefficient, 2 real-number multiplications for multiplying data (second term in Eq.~\eqref{iirfilter}) with a complex  normalization coefficient to yield proper SNR output, 2 real-number additions for combining the previous two products, while finally 2 real-number additions for adding the result of this filter into the total output.   

If we ignore costs for down- and up-sampling, which are performed relatively rarely, the total computational cost for initial-LIGO filters in Table \ref{tab_downsample} is
\begin{equation}\label{CCiir}
  \mathcal{C}_{\mbox{\scriptsize{IIR}}}=\sum_{k=0}^5 12S_k N_{\mathrm{\scriptsize{IIR,k}}}\simeq 2.4 \times 10^6 \mbox{flops}.
\end{equation}


On the other hand, if we carry out the same down sampling scheme for FIR filtering, the number of points in group $0$ will be
\begin{equation}
  N_{\rm FIR,0} = S_0 \cdot \left [ t(64\,\mathrm{Hz}) - t_{\rm ini}\right]
\end{equation}
where $t(64\,\mathrm{Hz})$ is the time at which the instantaneous frequency is 64\,Hz. For $k=1,2,3,\ldots 5$, we have
\begin{equation}\label{NFIR}
  N_{\rm FIR,k} = S_k \cdot \left [ t(2^{k+6}\,\mathrm{Hz}) -t(2^{k+5}\,\mathrm{Hz})\right]
\end{equation}
At sample rate $S_k$, for each time step, we have to perform two real-valued  correlations with array length $N_{\rm FIR\,k}$, which cost $4N_{\rm FIR,k}$ floating point operations. The total computational cost of FIR filtering is therefore 
\begin{equation}\label{CCfir}
  \mathcal{C}_{\mbox{\scriptsize{FIR}}}=\sum_{k=0}^{5} 4 S_k N_{\mathrm{\scriptsize{FIR,k}}}\simeq 2.0\times 10^7 \mbox{flops}.
\end{equation}
This is nearly $8$ times the cost of the IIR filter method assuming downsampling technique applied to both filtering methods.  The result of above cost estimation for the IIR and FIR filtering are also listed in Table~\ref{tab_downsample}.  We will show in the next subsections that the improvement is much more significant for advanced detectors as they venture into lower frequencies.

\subsection{ Dependence on initial frequency and future detectors\label{CCscaling}}

As initial frequency $f_{\rm min}$ is lowered in future gravitational-wave detectors, we anticipate much longer signals (see Table~\ref{basic_info}), and therefore a possibly dramatic increase of computational cost.  In this subsection, we will first obtain  analytical scalings in IIR and FIR computational costs, assuming an idealized down-sampling scheme.  We will then provide more realistic estimates of cost by constructing actual IIR filters and adopting the same successive $2\times$ down-sampling strategy. 

\begin{figure}
\includegraphics[width=0.45\textwidth]{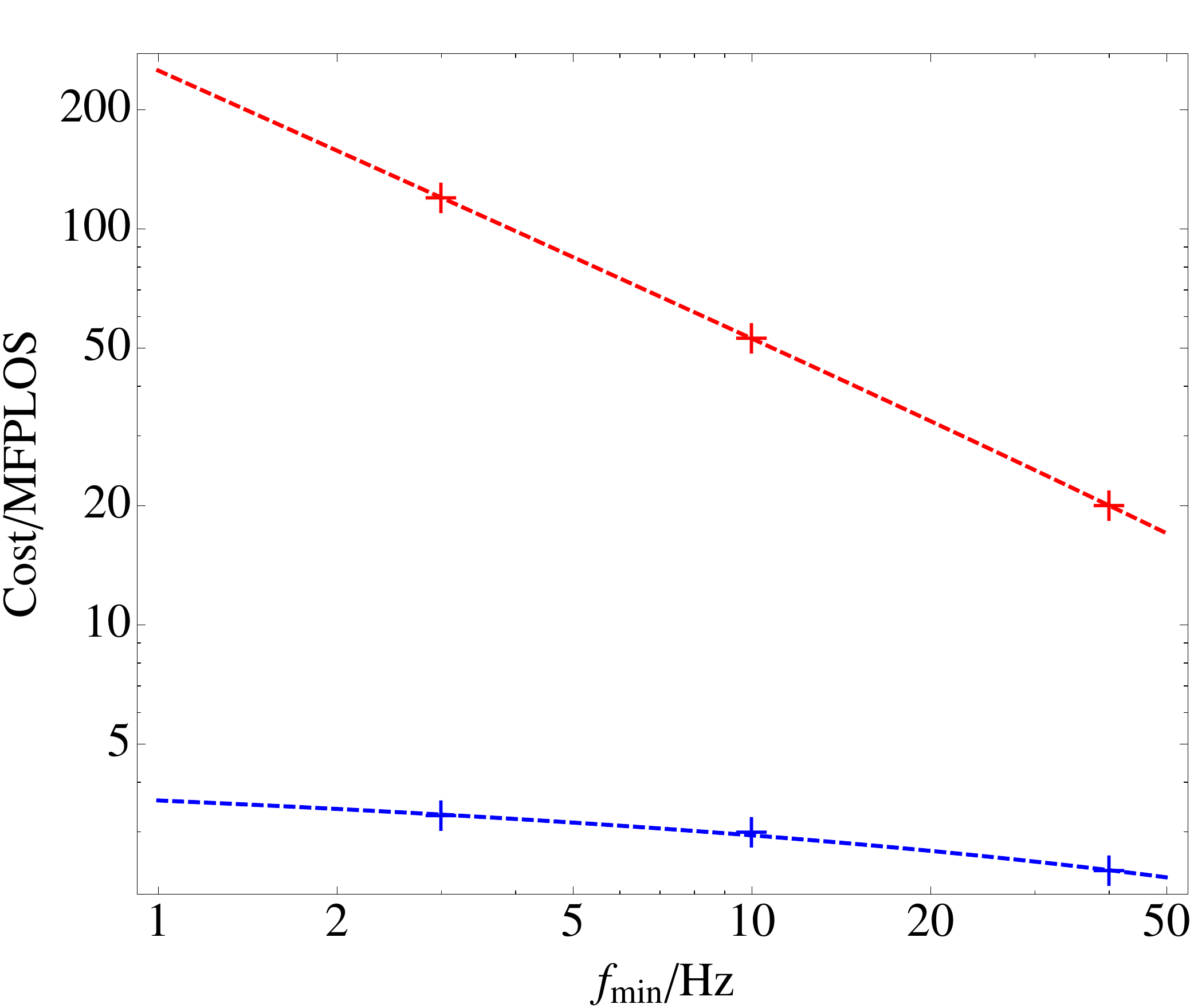}
\caption{Theoretical (dashed curves) and {numerical} (labeled by ``+"'s)  scaling of the computational cost with $f_{\rm min}$ for the FIR (red color) and IIR  (blue color)  method for one template, fixing $f_{\rm max}=2000\,$Hz.  The theoretical scaling is based on Eqs.~\eqref{CCfir2} and \eqref{CCiir2} (see Sec.~\ref{subsec:NE}), numerical values are taken from Table~ \ref{tab_downsample}, column 15. \label{fig:scaling}} 
\end{figure}

\subsubsection{Analytical Estimates}

Ideally, the minimum sample rate is twice the instantaneous frequency of the signal, or $S =2 f$.  For FIR filters, we have
\begin{eqnarray}
  N_{\mbox{\scriptsize{FIR}}}&\ge& 2N_{\mathrm{cyc}}\sim \int_{t_{\mathrm{ini}}}^{t_c} \Omega dt =\int\frac{\Omega}{\dot\Omega}  d\Omega \sim  f^{-5/3}_{\mathrm{min}}\,.
\end{eqnarray}
Converting the summation Eq.~\eqref{CCfir} into integral, we obtain:
\begin{equation}\label{CCfir2}
  \mathcal{C}_{\mbox{\scriptsize{FIR}}} \sim \int \Omega dN_{\mbox{\scriptsize{FIR}}}\sim f_{\mathrm{min}}^{-2/3}.
\end{equation}

For IIR filters, during  a dephasing time of $T=\sqrt{2\epsilon/\dot\Omega}$, we use one filter $\Delta N_{\mbox{\scriptsize{IIR}}}=1$ and $\Delta \Omega=\dot\Omega T=\sqrt{2\epsilon \dot\Omega}$, which leads to
\begin{equation}
  N_{\mbox{\scriptsize{IIR}}}=\int d\Omega \frac{dN_{\mbox{\scriptsize{IIR}}}}{d\Omega}=\int \frac{d\Omega}{\sqrt{2\epsilon \dot\Omega}}\sim 
  f_{\mathrm{min}}^{-5/6}.
\end{equation}
The computational cost of IIR filtering is
\begin{equation}\label{CCiir2}
  \mathcal{C}_{\mbox{\scriptsize{IIR}}}\sim\int \Omega dN_{\mbox{\scriptsize{IIR}}}\sim f_{\mathrm{max}}^{1/6}-f_{\mathrm{min}}^{1/6}.
\end{equation}
Note that for IIR filtering, the positive power law means the computational cost scales predominantly with the higher cut-off frequency, instead of the lower cut-off frequency --- we therefore expect the computational cost not to increase dramatically when $f_{\rm min}$ is lowered, if we already have $f_{\rm max} \gg f_{\rm min}$.

\subsubsection{Numerical Estimates} \label{subsec:NE}

More detailed constructions for Advanced LIGO and Einstein Telescope (ET$_{\rm }$) have been carried out, following Sec.~\ref{subsec:implementation}, assuming $f_{\rm min} = 10\,$Hz for Advanced LIGO and 3\,Hz for ET.  Assuming the same successive $2\times$ down-sampling strategy, we evaluate the single template computational cost for $(1.4+1.4)M_\odot$ binaries for both FIR and IIR filtering.     As it turns out, using the same $\epsilon=0.269$, but $(\alpha,\zeta)=(2.5,4.25)$ for Advanced LIGO and $(\alpha,\zeta)=(2.25,4.5)$ for ET, will still give us match above 0.99. 

The number of filters in each down-sampling band, as well as computational cost break-down for a single template are shown in the second and third tiers of Table \ref{tab_downsample}, for Advanced LIGO and ET, respectively.  We also compare our numerical values with scaling laws predicted in Eqs.~\eqref{CCfir2} and \eqref{CCiir2}, which are plotted in dashed curves in Fig.~\ref{fig:scaling}.  [We determined the normalization of the theoretical formulas using  numerical values of computational cost at $f_{\rm min}= 40$\,Hz.] The agreement is remarkable, especially considering that our successive 2-fold down-sampling is not continuous, and therefore rather non-ideal.

As we can see from Table~\ref{tab_downsample} and Fig.~\ref{fig:scaling}, the IIR reduces computational cost from (multi-rate) FIR filtering by factor of 8 for initial LIGO.  As we move to lower starting frequencies, the saving factor increases to 18 and 40, respectively.  The single-template cost, even when we {\it extrapolate} $f_{\rm min}$ to the rather unlikely 1\,Hz, stays at several MFLOPS. 



\section{Interpolation between IIR filters of different inspiral waveforms\label{filtersubgroups}}

In order to search for all possible kinds of compact binary coalescence, we must match the signal with a {\it family of templates} parametrized continuously by the parameters of the binary, e.g., their masses.  In practice, although maximization of match over certain parameters (e.g., orbital phase of the binary) can be done analytically, for the rest of the parameters, we must sample them discretely, and build a template bank --- and match the signal with each member of the bank.  The density of the discretization is usually determined by imposing that each member of the continuous family can be approximated well enough by at least one member of the bank, with mismatch less than a maximum tolerable value, $\varepsilon_{\rm max}$.  

For advanced detectors, the number of templates can be as large as $10^5$ \cite{Owen_Sathyaprakash_1999} posing a significant computational challenges.  Interpolation strategies have therefore been conceived (e.g., \cite{pinto00.1, pinto00.2, sanjeev, field11}) to {\it reduce} the number of templates, based on the fact that signal-to-noise ratio is a continuous function of the parameters being searched over.  More specifically, if we refer to the bank constructed by imposing the maximum tolerance of mismatch $\varepsilon_{\rm max}$ as the {\it fine bank}, then the hope is that even if match is {\it calculated} for a {\it coarse bank} in which parameters are less densely populated, the signal-to-noise ratio of the fine bank can still be {\it recovered by interpolation}, in such a way that the total cost of {\it computing} coarse-bank SNRs plus {\it interpolating} fine-bank SNRs is less than the cost of directly computing fine-bank SNRs.

Our interpolation method differs from previous work in that we divide each coarse-bank template into several sub-templates in frequency (thus time) domain, and recover fine-bank SNRs using SNRs from the sub-templates.  This approach has been inspired by the SVD approach~\cite{svd, svd2} adopted by the  LLOID~\cite{LLOID} and the 2-bank interpolation in MBTA~\cite{MBTA} methods.  We will show that, although the division into sub-templates increases the cost of recombination, it allows a much coarser bank --- and finally decreases the computational cost by a large factor.



\subsection{Template banks in general}
\label{subsec:tplbank}

To develop a scheme to discretize the parameter space without losing detection efficiency, we must know how much the SNR is reduced by using a template whose parameter values differ from those of the signal. We define the mismatch between two normalized templates of different sets of  parameters as
\begin{equation}\label{mismatch}
  \varepsilon \equiv  1- \langle u(\bm{\lambda})|u(\bm{\lambda'})\rangle.
\end{equation}
The template $u$ is specified by a parameter vector $\bm{\lambda}$. If $\bm{\lambda'}$ is near to $\bm{\lambda}$, we can Taylor expand $\varepsilon$ at $\bm{\lambda}$ and have the approximation to second order of $\Delta\bm{\lambda}\equiv \bm{\lambda'}-\bm{\lambda}$ as
\begin{equation}
  \label{metricdefinition1}
  \varepsilon\simeq \frac 1 2 \frac {\partial^2\varepsilon} {\partial \lambda_i\partial\lambda_j}\bigg|_{\Delta\bm{\lambda}=0} \Delta\lambda_i\Delta\lambda_j,
\end{equation}
from which we define a (positive definite) metric in the parameter space
\begin{equation}
  \label{metricdefinition2}
  \gamma_{ij}\equiv \frac 1 2\frac {\partial^2\varepsilon} {\partial \lambda_i\partial\lambda_j}\bigg|_{\Delta\bm{\lambda}=0}.
\end{equation}
Equations~\eqref{metricdefinition1} and \eqref{metricdefinition2} indicates that mismatch between neighboring points in the parameter space can be viewed as distance measured by metric {\boldmath{$\gamma$}}. 

Suppose we would like to place a template bank in a $D$-dimensional parameter space, with a mismatch no higher than $\varepsilon$, then the most straightforward strategy would be laying down a cubic grid with {\it proper side length} $dl$ measured by the metric $\gamma_{ij}$, such that template placed at each grid point will  be able to cover a cube whose vertices are centers of neighboring cubes.  This means we have
\begin{equation}
  D(dl/2)^2 =\varepsilon\,.
\end{equation}
The volume spanned by each cube  (according to metric $\gamma_{ij}$) is therefore
\begin{equation}
  \Delta V = dl^D = (2\sqrt{\varepsilon/D})^D.
\end{equation}
The total number of templates in the bank would be the total volume of the parameter space divided by the volume of each cell, or
\begin{equation}
  \label{Ntemplate}
  \mathcal{N} =\frac{V_{\rm tot}}{\Delta V}=\frac{\int d^D \bm{\lambda} \sqrt{det \| \gamma_{ij}\|}}{(2 \sqrt{\varepsilon/D})^D}
\end{equation}

\subsection{Newtonian Chirps}

\label{templatebank}

Through the Stationary-Phase Approximation~\cite{SPA}, the Fourier Transform of a Newtonian Chirp can be written as
\begin{equation}\label{newtonian frequency}
  \tilde u(f;A,t_c,\phi_c) \propto  f^{-7/6} e^{i(A f^{-5/3}+2\pi f t_c+\phi_c)},\; f>0\,,
\end{equation}
and  $\tilde u(f) = \tilde u^*(-f)$ for $f<0$. The mismatch between two neighboring templates with parameters ($A$, $t_c$, $\phi_0$) and ($A+\Delta A$, $t_c+\Delta t_c$, $\phi_c+\Delta \phi_0$) can be written as
\begin{equation}
  \label{mismatchChirp}
  \varepsilon (\Delta A, \Delta t_c, \Delta \phi_0) = 1- \frac
  {\displaystyle \int_{f_{\rm min}}^{f_{\rm max}}\frac{f^{-7/3} \cos \Delta\Phi}{S_h(f)}df}
  {\displaystyle \int_{f_{\rm min}}^{f_{\rm max}}\frac{f^{-7/3}}{S_h(f)}df}
\end{equation}
where
\begin{equation}
  \Delta\Phi =f^{-5/3} \Delta A+2\pi f\Delta t_c +\Delta \phi_c
\end{equation}
Expanding Eq.~\eqref{mismatchChirp} up to second order in $\Delta\Phi$, we obtain by comparing with Eqs.~\eqref{metricdefinition1} and \eqref{metricdefinition2} the metric
\begin{equation}
\label{gamma}
  \|\gamma_{ij}\|=
  \left[ \begin{array}{ccc}
      I(-\frac{17}{3})& I(-3) & I(-4)\\
      *& I(-\frac 1 3) &  I(-\frac 4 3)\\
      * & * & I(-\frac 7 3)
    \end{array} \right],
\end{equation}
where ``$*$'' indicates terms obtainable by symmetry, and 
\begin{equation}
  I(\beta)=\frac{1}{2}\left[\int_{f_{\mathrm{min}}}^{f_{\rm max}} df \frac{f^{\beta}}{S_h(f)}\right]\bigg/\left[\int_{f_{\mathrm{min}}}^{f_{\rm max}}df \frac{f^{-7/3}}{S_h(f)}\right],
\end{equation}
and we have used $i=1,2,3$ to label $\Delta A$, $2\pi\Delta t_c$ and $\Delta \phi_c$, respectively.  Note the metric depends on the frequency division and noise spectral density only.

Here among the three parameters, search over $\phi_c$ is done analytically, as discussed in  Sec.~\ref{sec:match}, while search over $t_c$ is carried out systematically at the sample rate --- the only parameter left to discretize is $A$.  Therefore,  $A$ is an intrinsic parameter as described previously.   The correct way to place templates along intrinsic parameter directions is to ``project out'' the intrinsic parameters, as discussed, e.g.,  by Owen and Sathyaprakash~\cite{Owen_Sathyaprakash_1999}.  

In our case, the projected metric along direction $A$ is one dimensional given by 
\begin{equation}
\label{project}
g_{11} = \gamma_{11} -\frac{\gamma_{13}^2\gamma_{22}-2\gamma_{12}\gamma_{13}\gamma_{23} +\gamma_{12}^2\gamma_{33}}{\gamma_{22}\gamma_{33}-\gamma_{23}^2}
\end{equation}
which depends on $f_{\mathrm{min}}$, $f_{\mathrm{max}}$  and the noise curve $S_h$ through $I(\beta)$.    
Following Eq.~\eqref{Ntemplate}, the  number of templates required to achieve a mismatch $\varepsilon$ is then
\begin{equation}
  \mathcal{N}=\frac{\sqrt{g_{11}}(A_{\rm max}-A_{\rm min})}{2\sqrt{\varepsilon}}, 
\label{Ntemplates}
\end{equation}
where $A_{\rm min}$ and $A_{\rm max}$ are the minimum and maximum values of $A$.    Here we can be more specific about template placement along the $A$ direction.   Given any $A$, which is associated with a member of the template bank, and suppose its mismatch with a neighboring template with $A \pm  \Delta A$ is $\varepsilon_{\rm max}$, or
\begin{equation}
\label{DAfine}
g_{11} (\Delta A)^2 =\varepsilon_{\rm max}
\end{equation}
then neighboring templates should be placed at $A \pm 2\Delta A$, therefore we have
\begin{equation}
\mathcal{N} =\frac{A_{\rm max}-A_{\rm min}}{2\Delta A}
\end{equation}
which recovers Eq.~\eqref{Ntemplates}.

Here we give the noise spectral density we use for initial LIGO, Advanced LIGO, and Einstein Telescope (${\rm ET}_{\rm B}$).   For the initial LIGO~\cite{SF_iLIGO}, we have $x=f/(150\,{\rm Hz})$ and 
  \begin{equation}
  \label{ShI}
    S_h(f)=9\cdot 10^{-46}\left[(4.49 x)^{-56}+0.16 x^{-4.52}+0.52+0.32x^2\right]. 
  \end{equation}
For Advanced LIGO~\cite{SF_aLIGO}, we have $x=f/(215\,{\rm Hz})$ and
  \begin{equation}
    S_h(f)=10^{-49}\left[ x^{-4.14}-5x^{-2}+111\frac{1-x^2+\frac{1}{2} x^4}{1+\frac{1}{2} x^2}\right].
  \end{equation}
Note this is different from what is used in \cite{LLOID}.  As a result, two methods are dealing with different number of templates for the same parameter space.  This should be taken into account when we compare the computational cost of the two methods.  For the Einstein Telescope~\cite{SF_ET}, we have $x=f/(100\,{\rm Hz})$ and 
  \begin{eqnarray}\label{ShETB}
    \sqrt{S_h(f)}=10^{-25}&\Big(&2.39\times 10^{-27} x ^{-15.64}+0.349 x^{-2.145}\nonumber \\ &+&1.76x^{-0.12}+0.409x^{1.10}\Big).
  \end{eqnarray}

Applying Eqs.~\eqref{Ntemplates} and \eqref{DAfine} to these three detectors, we can show that the number of templates increase by a factor of 3.9 when we upgrade from initial to Advanced LIGO, and another factor of 4.4 when we upgrade from Advanced LIGO to the Einstein Telescope.  These numbers are listed in Table~\ref{basic_info}, column 5. 

\subsection{Subtemplates}
\label{subtemplate}

\subsubsection{General Discussion}

Now suppose we divide our entire signal frequency interval, $(f_{\rm min},f_{\rm max})$ into $M$ segments of 
\begin{equation}
[f_0,f_1],\; [f_1,f_2], \; \ldots\;, [f_{M-1},f_{M}],
\end{equation}
with $f_0=f_{\rm min}$ and $f_{n}=f_{\rm max}$. (When we later apply this to IIR filter chains, $M$ will be much less than the total number of filters, $N$.)  For any template $u$, we define sub-template $u_J$, $J=1,\ldots M$, {to have the same value as template $u$ within the frequency interval $[f_{J-1},f_J]$ but have zero values elsewhere,}
\begin{equation}
  \tilde u_J (f)=\left\{\begin{array}{cl} \tilde u(f),  & f_{J-1}\le f \le f_{J}, \\
      \\
      0, & \mbox{otherwise}.
    \end{array}
  \right.
  \label{fj}
\end{equation}   
Now let us consider two neighboring templates, $u$ and $v$, their $J^{th}$-{\it sub-innerproduct} can be naturally defined as an integral over frequency segment $J$:
\begin{equation}
  \langle u|v \rangle_J\equiv \langle u_J | v_J\rangle =4 \mathrm{Re}\left[ \int_{f_{J-1}}^{f_{J}} df \frac{\tilde u^*(f) \tilde v(f)}{S_h(f)}\right]\,.
\end{equation}
This sub-innerproduct can also be regarded as the contribution to the full inner product $\langle u | v\rangle$ from segment $J$ [Cf.~Eq.~\eqref{innerproduct}], and
\begin{equation}
  \label{subinner}
  \langle u | v\rangle = \sum_{J=1}^{M} \langle u|v \rangle_J
\end{equation}

We denote $u$ and $u+\Delta u$ as neighboring templates, and we also define their {\it $J^{\rm th}$-sub-mismatch} specific to interval $J$, in the intrinsic parameter space, as
\begin{equation}
  \varepsilon_J\equiv  1-\frac{\langle u|u+\Delta u\rangle_J}{\sqrt{\langle u|u\rangle_J\langle u+\Delta u|u+\Delta u\rangle_J}},
\end{equation}
which is equal to the ``ordinary" mismatch between $u_J$ and $u_J+\Delta u_J$ as defined in Eq. (\ref{mismatch}).  Up to second order in $\Delta u$, we can show that the total mismatch and the
$J^{\rm th}$-sub-mismatch are 
\begin{eqnarray}
  \varepsilon&=&\frac 1 2 \frac{\langle \Delta u|\Delta u\rangle}{\langle u|u\rangle}\\
  \varepsilon_J&=& \frac 1 2 \frac{\langle \Delta u|\Delta u\rangle_J}{\langle u|u\rangle_J}
\end{eqnarray}
Using Eq.~\eqref{subinner}, we can show that
\begin{equation}\label{weightedmean}
  \varepsilon=\sum_{J=1}^M \varepsilon_J \frac{\langle u|u\rangle_J}{\langle u|u\rangle}.
\end{equation}
Since
\begin{equation}
  \sum_{J=1}^M\frac {\langle u|u\rangle_J}{\langle u|u\rangle}=1,
\end{equation}
the overall mismatch is therefore a weighted average of the sub-mismatches. This means to achieve an overall mismatch of $\varepsilon$, we  only need to make sure the sub-mismatches $\varepsilon_J$ average to $\varepsilon$.  This has dramatic implications in the sense that it allows the overall mismatch to be maintained by (1) dividing the frequency band into  several frequency intervals with non-uniform sub-mismatches,  (2) reducing the size of frequency intervals to allow larger step size for intrinsic parameters. These lay the foundation for our template interpolation method.


\begin{figure}[htbp]
  \begin{center}
    \includegraphics[width=3in]{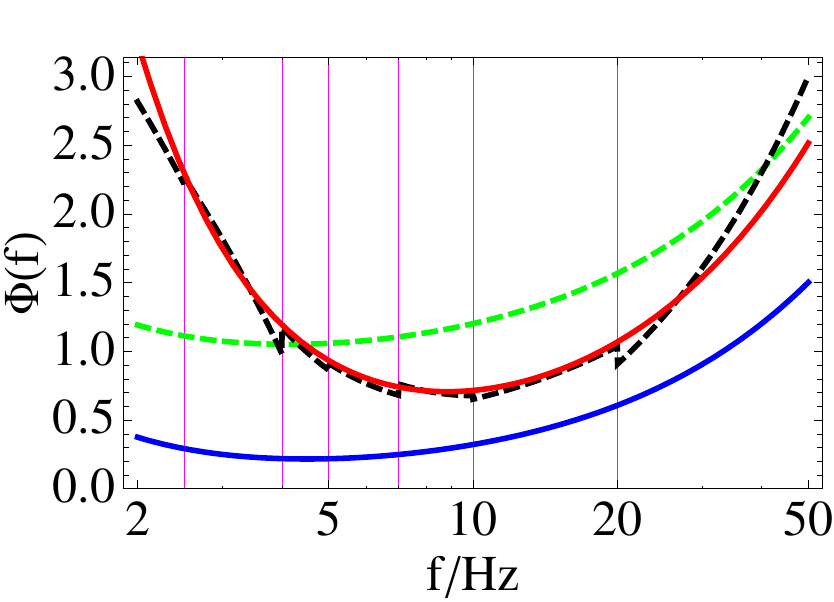}
  \caption{{Illustration of the phase function $\Phi (f)$ vs frequency for the presumed parameter $A$ (blue solid line) and its neighboring parameter $A+\Delta A$ (red line), the linear shift of the blue line to match the red line (green dashed line),  and a piecewise approximation (black dashed line) of the red line by shifting segments from the blue line.}   It shows  that with smaller frequency intervals, it is easier to match phases arising from different intrinsic parameters. \label{frequencyseg}}
  \end{center}
\end{figure}

To qualitatively understand the reason that {the grid size for intrinsic parameter placement} can be enlarged when we restrict ourselves to smaller frequency intervals, we first note that in the frequency domain, it is the phase that we need to match, while the amplitude as a function of frequency is the same for all parameters.  We note that the phase of $\tilde u(f)$, which we denote by $\Phi(f)$, is determined by $A$, as well as $t_c$ and $\phi_c$ (Eq.~(\ref{newtonian frequency})).   In Fig.~\ref{frequencyseg}, we plot the phase $\Phi(f)$ for a particular set of parameters $(A,t_c,\phi_c)$ in blue and also for a neighboring set of parameters $(A+\Delta A,t_c,\phi_c)$ in red. If we were to use the template with parameter $A$ to search for a signal with parameter $A+\Delta A$, we could shift $\phi_c$ and $t_c$ used in the search, which corresponds to shifting the blue curve by a linear function in frequency.  The green dashed line illustrates a reasonably optimal attempt --- yet the difference between the green curve and the red curve cannot be reconciled very well due to the fact that linear functions do not correct for curvature. However, if we divide the frequency range into several intervals, and allow different values of $\Delta t_c$ and $\Delta \phi_c$ to be applied to each interval, then sub-templates with $A$ can achieve rather low sub-mismatches with signal with $A+\Delta A$. This corresponds to the fact that a curve can be better approximated by straight lines when divided into smaller intervals.




\subsubsection{Newtonian Chirp in the Frequency Domain}
\label{NCintp}

Let us now focus on a particular frequency segment $J$, with $f_{J-1} \le f \le f_J$, and work out the relation between $\Delta A$ and $\varepsilon_J$, as $\Delta\phi_c$ and $\Delta t_c$ are allowed to readjust their values (to be different from other segments).  This simply requires us to repeat the procedure in Sec.~\ref{templatebank} for each segment: with $\Delta A$, $\Delta t_c$ and $\Delta \phi_c$, we have the $J^{\rm th}$-sub-mismatch of 
\begin{equation}
\label{submistmach}
\varepsilon_J = 
\left[\begin{array}{ccc}\Delta A &2\pi\Delta t_c & \Delta\phi_c\end{array}\right]
{\mbox{\boldmath{$\gamma$}}}^{J}\left[\begin{array}{ccc}\Delta A\\2\pi\Delta t_c\\\Delta \phi_c\end{array}\right]
\end{equation}
with
\begin{equation}
{\mbox{\boldmath{$\gamma$}}}^{J}\equiv\left[\begin{array}{ccc}
      I_J(-\frac{17}{3})&I_J(-3)&I_j(-4)\\
      *&I_J(-\frac 1 3) & I_j(-\frac 4 3)\\
      *& *& I_J(-\frac 7 3)
    \end{array}\right]\,.
\end{equation}
and
\begin{equation}
  I_J(\beta)\equiv \frac{1}{2}\left[\int_{f_{J-1}}^{f_{J}} \frac{df f^{\beta}}{S_h(f)}\right]\Big/\left[\int_{f_{J-1}}^{f_{J}}\frac {df\ f^{-7/3}}{S_h(f)}\right].
\end{equation}
Note that the above are identical to Eqs.~\eqref{mismatchChirp}--\eqref{gamma}, except with integrations restricted to the interval of $[f_{J-1},f_{J}]$. 

The next step is similar to the ``projection'' process described by Owen and Sathyaprakash, but restricted to interval $J$. With Eq.~\eqref{submistmach}, we ask the following question: if we are allowed to freely re-adjust individually the values of  $\Delta t_c$ and $\Delta \phi_c$ for  interval $J$ of the template (i.e., the $J^{\rm th}$-subtemplate), what would be the $J^{\rm th}$-sub-mismatch achievable for $\Delta A$, and what should the corresponding $\Delta\phi_c$ and $\Delta t_c$ be.

The answer to the question is readily obtainable by a maximization of the mismatch $\varepsilon$ over $\Delta t_c$ and $\Delta \phi_c$, fixing $\Delta A$. This results in adjustments of
\begin{equation}
\label{correction}
\left[
\begin{array}{c}
2\pi \Delta t_c^J \\
\Delta \phi_c^J
\end{array}\right]
=-\left[
\begin{array}{cc}
\gamma_{22}^{J} & \gamma_{23}^{J} \\
\gamma_{32}^{J} & \gamma_{33}^{J} 
\end{array}
\right]^{-1} 
\left[
\begin{array}{c}
\gamma_{12}^{J} \\
\gamma_{13}^{J} 
\end{array}
\right]\Delta A
\end{equation}
which result in the $J^{th}$ sub-mismatch of
\begin{equation}
\varepsilon_J =g_{11}^{J} (\Delta A)^2 \,,
\end{equation}
with
\begin{equation}
g_{11}^J \equiv   \gamma_{11}^J 
-
\frac{[\gamma_{13}^{J}]^2\gamma_{22}^{J}-2\gamma_{12}^{J}\gamma_{13}^{J}\gamma_{23}^{J} +[\gamma_{12}^{J}]^2\gamma_{33}^{J}}
{\gamma_{22}^{J}\gamma_{33}^{J}-[\gamma_{23}^{J}]^2}.
\end{equation}
{Following Eq.~\eqref{weightedmean}, we have the total mismatch}
\begin{equation}
\varepsilon =  {g_{11}^{\rm eff}} (\Delta A)^2\,, 
\end{equation}
where 
\begin{equation}
g_{11}^{\rm eff} = \sum_J \frac{g_{11}^{J}\langle u | u\rangle_J}{\langle u | u\rangle}  
\end{equation}
is an {\it effective metric coefficient} for any division of the frequency band. More specifically, $g_{11}^{\rm eff}$ describes the mismatch achievable by individually adjusting $\Delta\phi_c^{J}$ and $\Delta t_c^{J}$, for each interval of the division.  It can be shown that in general a finer division of frequency intervals always gives a smaller $g_{11}^{\rm eff}$.   As a consequence, if we define 
\begin{equation}
\Delta A_{\rm cb} =\sqrt{\frac{\varepsilon_{\rm max}}{g_{11}^{\rm eff}}}\,,
\label{DAcoarse}
\end{equation}
with the subscript ``cb'' indicating coarse bank, then $\Delta A_{\rm cb}$ will be greater than $\Delta A$ given by Eq.~\eqref{DAfine} where $g_{11}$ is evaluated using the full frequency band. In order to make a distinction, we shall rewrite that same equation as 
\begin{equation}
\Delta A_{\rm fb}=\sqrt{\frac{\varepsilon_{\rm max}}{g_{11}}}\,,
\end{equation}
but adding a subscript ``fb'' to indicate the fine bank.  In order to maximize $\Delta A_{\rm cb}$ for a maximum mismatch $\varepsilon_{\rm max}$,  we should simply minimize $g_{11}^{\rm eff}$ globally, over all possible frequency division schemes.   Because a template at $A$ in the fine bank covers $(A-\Delta A_{\rm fb}, A+\Delta A_{\rm fb})$, the ratio of the number of templates in the coarse bank to that of the fine bank is,
\begin{equation}
\frac{\mathcal{N}_{\rm cb}}{\mathcal{N}_{\rm fb}} = \sqrt{\frac{g_{11}^{\rm eff}}{g_{11}}}
\end{equation}

In summary, given a required maximum mismatch $\varepsilon_{\rm max}$ with a particular frequency subdivision, by adjusting $\Delta \phi_c^J$ and $\Delta t_c^J$ individually,  a single template at $A$ can cover the region $(A-\Delta A_{\rm cb}, A+\Delta A_{\rm cb})$.  For a signal with $|\Delta A| \le \Delta A_{\rm cb}$, the $J^{th}$-sub-template {for parameter $A+\Delta A$ can} be constructed by adjusting $\Delta t_c^{J}$ and $\Delta\phi_c^{J}$ of the sub-template of template $A$ using Eq.~\eqref{correction}.   The {interpolated template of parameter $A+\Delta A$  is therefore the sum of the constructed sub-templates from a coarse-bank template $A$}
\begin{eqnarray}\label{interpolation1}
  &&\tilde u(f;A+\Delta A,t_c,\phi_c)\nonumber\\
  &&=\sum_{J=1}^{M}\tilde u_j(f;A,t_c+\Delta t_{c}^J(\Delta A),\phi_c+\Delta\phi_{c}^J(\Delta A))\nonumber\\
  &&=\sum_{J=1}^{M}\tilde u_j(f;A,t_c,\phi_c) e^{i 2\pi f \Delta t_{c}^J(\Delta A)+i\Delta\phi_{c}^J(\Delta A)}.
\end{eqnarray}

It is straightforward to establish the following properties of the effective metric: (i) $g_{11}^{\rm eff}$ always becomes smaller when we insert one or more dividing frequencies into an existing division of $[f_{\rm min},f_{\rm max}]$, (ii) if we continue to decrease the maximum size of intervals, we can decrease $g_{11}^{\rm eff}$ indefinitely [in fact, for small intervals,  $g^{J}_{11}$ scales as $(\Delta f)^5$, which means $g_{11}^{\rm eff}$ should scale as $(\Delta f)^4$, and hence $\Delta A$ scales as $(\Delta f)^{-2}$].  Furthermore, for template families with {\it more than one parameter}, it is straightforward to generalize our result to 
\begin{equation}
g_{ab}^{\rm eff} = \sum_J \frac{g_{ab}^J \langle u | u \rangle_J }{\langle u | u\rangle}
\end{equation}
with the number of templates in the coarse bank given by 
\begin{equation}
\frac{\mathcal{N}_{\rm cb}}{\mathcal{N}_{\rm fb}} = \sqrt{  \frac{\mathrm{det}\| g_{ab}^{\rm eff} \|}{\mathrm{det} \| g_{ab}\|}}
\end{equation}

\begin{figure*}[htbp]  
\includegraphics[width=4.57in,angle=-90]{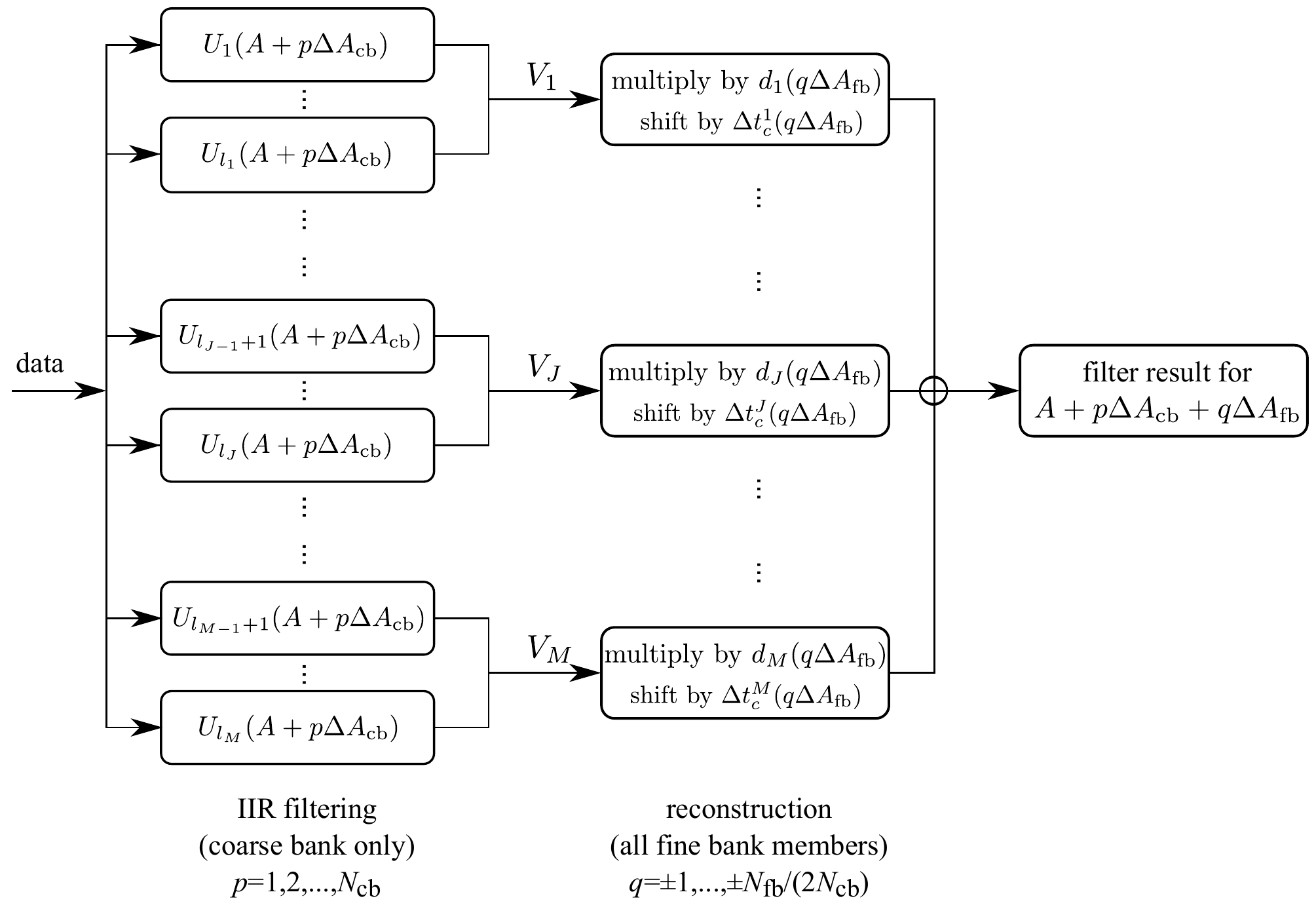} 
\caption{Schematic diagram of the IIR filtering process for a template with parameter $A+p\Delta A_{\rm cb}+q \Delta A_{\rm fb}$. The first part is the IIR filtering for a member of the coarse bank, $A+p\Delta A_{\rm cb}$, which produces a range of filter outputs, labeled by $U_1$ \ldots $U_{l_M}$.  These are grouped into $M$ groups of summed IIR results $V_1$, \ldots, $V_M$. The result for $A+ p\Delta_{\rm cb}+q\Delta A_{\rm fb}$ is obtained by combining these $V_J$'s  after each one is  multiplied by $d_J(q\Delta A_{\rm fb})$ and shifted by $\Delta t_c^J(q\Delta A_{\rm fb})$. The entire data analysis process still computes $N_{\rm fb}$ filter results, by including $N_{\rm cb}$ possible $p$'s and $N_{\rm fb}/N_{\rm cb}$ possible $q$'s for each $p$.  [In the special case of $q=0$, the $V_J$'s are directly summed without having to go through multiplications and time shifts.]  The downsampling or upsampling process is not shown.  \label{timeshift}} 
\end{figure*}


\subsection{Application to IIR filtering technique}\label{IIRintp}

In this section, we will apply the formalism developed in the previous subsection and  discuss how we can implement IIR filter chains only for a much coarser bank of templates --- while still obtaining SNRs for the entire fine template bank.  Discussions made in the previous sections, although strictly speaking only apply to sharp divisions in the signal frequency band, still qualitatively apply to IIR filters that work in time-domain.   The trick is to replace {\it frequency intervals} in the previous section by {\it groups of IIR filters}. This approach will work as long as we include enough number of filters in each ``group'', so that overlaps between different groups are relatively unimportant.  We note that, as is the case for the construction of IIR filter chains, the construction of the interpolation scheme by itself does not justify its efficiency --- a separate test of achievable match will be carried out explicitly after the interpolation scheme is constructed.

\begin{figure*}
\includegraphics[width=5.5in]{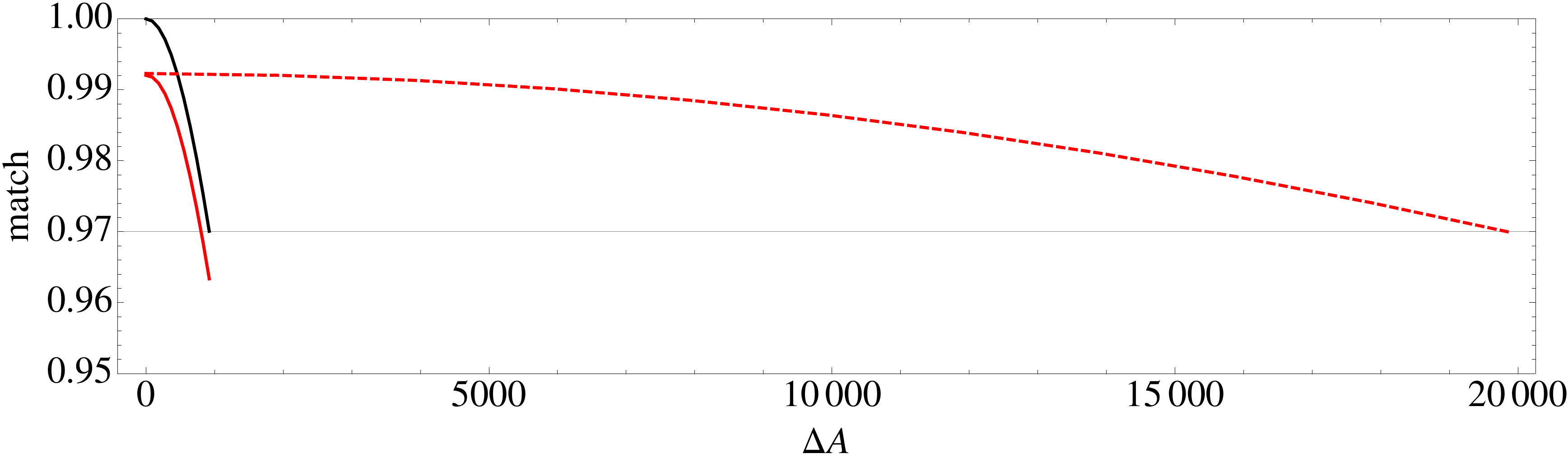} 
\caption{Matches achievable with a Newtonian-Chirp signal at $A+\Delta A$, by various templates built for $A$, using initial LIGO noise spectral density {for a (1.4+1.4)\msun\  NS-NS binary}.  Black solid curves corresponds to the result for a Newtonian-Chirp template, therefore the match is equal to unity at $\Delta A=0$.  Red solid curve corresponds to that using IIR filters, while red dashed curve corresponds to the interpolated match that can be recovered by using 6 filter subgroups.} 
\label{ovp}
\end{figure*}

To be more specific, we re-group the entire chain of $N$ IIR filters  into $m$ sub-groups, with group $J$ including those whose oscillation frequency lies within the frequency interval $J$ defined in Sec.~\ref{subtemplate}.  In other words, group $J$ of IIR filters can be written as
\begin{eqnarray}
&&V_J(t;A,t_c,\phi_c) \nonumber\\
&=& \sum_{\frac{\Omega_l}{2\pi} \in [f_{J-1},f_J]} U_l (t;A,t_c,\phi_c) \nonumber\\
&\equiv &\sum_{l=l_{J-1}+1}^{l_{J}} U_l(t;A,t_c,\phi_c) ,\quad J=1,\ldots,M\,,
\end{eqnarray} 
where we have $l_0 =0$.  We will treat $V_J$ as corresponding to the  $\tilde u_J(f;A,t_c,\phi_c)$ of Sec.~\ref{subtemplate}.  As a consequence, from Eqs.~\eqref{correction} and \eqref{interpolation1},  
signal $u(t;A+\Delta A,t_c,\phi_c)$ can be interpolated by the IIR filters constructed for $u(t;A,t_c,\phi_c)$
\begin{eqnarray}\label{interpolation3}
  &&u(t;A+\Delta A,t_c,\phi_c)
 \nonumber\\
  &\simeq&
   \sum_{J=1}^{M}  V_J(t;A,t_c+\Delta t_c^J,\phi_c+\Delta\phi_c^J) \nonumber\\
   &=&
   \sum_{J=1}^{M}e^{i\Delta\phi_c^J}  V_J(t;A,t_c+\Delta t_c^J,\phi_c)\,.
\end{eqnarray}
Here $\Delta t_c^J$ and $\Delta\phi_c^J$ should be computed from $\Delta A$ using Eq.~\eqref{correction}.

In practice we can easily generalize the coefficients in front of $V_J$ to further reduce the overall mismatch, by using a slightly more general reconstruction formula:
\begin{eqnarray}\label{interpolation5}
  u(t;A+\Delta A,t_c,\phi_c)\simeq
  \sum_{J=1}^{m} d_J V_J(t;A,t_c+\Delta t_c^J,\phi_c),
\label{generalsum}
\end{eqnarray}
where $d_J$ are complex coefficients that depend on $\Delta A$, given by
\begin{equation}
d_J = \sum_K T^{-1}_{JK} \langle V_K(A,t_c,\phi_c) |u(A+\Delta A,t_c,\phi_c)\rangle
\end{equation}
with the matrix $\mathbf{T}$ given by
\begin{equation}
T_{JK} = \langle V_J(A,t_c,\phi_c) | V_K(A,t_c,\phi_c) \rangle 
\end{equation}

\subsection{Full computational cost\label{Full computational cost}}

Fig.~\ref{timeshift} illustrates the procedure of obtaining the outputs from IIR filter chain for fine-bank coverage by interpolating  coarse-bank filter outputs described previously.  Upon obtaining outputs from subgroups of IIR filters for the {\it coarse bank}, we need to 
reconstruct outputs for all members of the {\it fine bank}. We hereby estimate the cost for reconstruction.  
Let's assume that a  member of the fine bank that is {\it not} a member for the coarse bank is $\Delta A$ away from a coarse-bank template $A$.  
For this $\Delta A$,   we need to go through each group $J$ of filters, take the total output of this group (which corresponds to filtering by $V_J$), multiply it by the complex number $d_J$  (6 floating-point operations) and shift in time by $\Delta t_c^J$, and then add it to the sum (2 floating-point operations).  The output eventually yields the SNR corresponding to the member of the fine bank. Note that both $d_J$ and $\Delta t_c^J$ are functions of $\Delta A$, but they do not need to be recalculated for each time step.

\begin{table*}[t]
  \begin{center}
    \begin{tabular}{|c||c|c|c|c|c|c|c|c|c|c|c|c|c|}
      \hline
 \multirow{2}{*}{}  & $S_k$ ($\mathrm{s}^{-1}$) & 16 & 32 & 64 &  128 & $256$ & $512$ & $1024$ & $2048$ & $4096$ &$8192$ & Cost\\
     \cline{2-1}
&  $f/\mbox{Hz}$ & {\footnotesize 2 -- 4} & {\footnotesize 4 -- 8} & {\footnotesize 8 -- 16}  &{\footnotesize 16 --  32} &{\footnotesize 32 -- 64} & {\footnotesize 64 -- 128}  &{\footnotesize 128 -- 256} & {\footnotesize 256 -- 512} & {\footnotesize 512 -- 1024} &{\footnotesize  1024 --  } & Total \\
      \hline\hline
      \multirow{3}{*}{iLIGO}  & 
      $N_{{\rm group},k}$ &   
       & 
       & 
       &
       & 
1      & 
2      & 
1      & 
1      &  
       &  
1      &  \multirow{3}{*}{0.10}\\   
      \cline{2-12}  
& 
      $\{f_J\} \cap (f_{k-1}, f_k]$ 
      &   
      & 
      & 
      & 
      & 
      {\footnotesize 52.9} & 
      {\footnotesize 71.0, 97.3}
      & 
      {\footnotesize 141}      & 
      {\footnotesize 244}      &  
      &  
      {\footnotesize 2000}  &  \\   
      \cline{2-12}  
      & $\mathcal{C}_{{\rm recomb},k}$ 
      & 
      & 
      & 
      & 
      & 
     {\footnotesize 0.002}  & 
       {\footnotesize 0.008} &
    {\footnotesize 0.008}   & 
     {\footnotesize 0.016}  &  
      &   
 {\footnotesize 0.066}    & \\
       \hline\hline
      \multirow{3}{*}{aLIGO}  & 
      $N_{{\rm group},k}$ &   
       & 
       & 
1       &
3       & 
2      & 
 2     & 
  1    & 
            &  
       &  
   1   &  \multirow{3}{*}{0.090}\\   
      \cline{2-12}  
& 
      $\{f_J\} \cap (f_{k-1}, f_k]$ 
      &   
      & 
  & 
{\footnotesize 12.9} 
        & 
\begin{tabular}{c}
{\footnotesize 16.8, 22.1} \\
{\footnotesize 29.6} 
\end{tabular}
& 
{\footnotesize 40.1, 55.2}
& 
      {\footnotesize 78.5, 122} & 
      {\footnotesize 228}      & 
             &  
      &  
      {\footnotesize 2000}  &  \\   
      \cline{2-12}  
      & $\mathcal{C}_{{\rm recomb},k}$ 
      & 
      & 
      & 
     {\footnotesize 0.0005} & 
     {\footnotesize 0.003}  & 
 {\footnotesize 0.004}      & 
       {\footnotesize 0.008} &
  {\footnotesize 0.008}     & 
       &  
      &   
 {\footnotesize 0.066}   &\\
       \hline\hline
      \multirow{3}{*}{ET$_{\rm B}$}  & 
      $N_{{\rm group},k}$ &   
       & 
2       & 
2       &
2       & 
1     & 
 1     & 
  1    & 
            &  
       &  
   1   &  \multirow{3}{*}{0.083}\\   
      \cline{2-12}  
& 
      $\{f_J\} \cap (f_{k-1}, f_k]$ 
       &   
  & 
{\footnotesize 5.1, 6.9}      & 
{\footnotesize 9.3,12.8}        & 
{\footnotesize 17.8, 25.3} & 
{\footnotesize 37.4} &  
{\footnotesize 60.3}&  
      {\footnotesize 122}& 
     & 
     &    
      {\footnotesize 2000}  &  \\   
      \cline{2-12}  
      & $\mathcal{C}_{{\rm recomb},k}$ 
      &  
      & 
      {\footnotesize 0.0005}& 
     {\footnotesize 0.001} & 
     {\footnotesize 0.002}  & 
 {\footnotesize 0.002}      & 
       {\footnotesize 0.012} &
  {\footnotesize 0.008}     & 
       &  
      &   
 {\footnotesize 0.066}   &\\
       \hline\hline
    \end{tabular}
    \caption{
    Break-down of recombination cost required for obtaining one fine-bank template using the interpolation method, for initial, Advanced LIGO and the Einstein Telescope --- assuming a successive two-fold down-sampling and ignoring the cost of down- and up-sampling.  The IIR filter information is listed in Table~\ref{tab_downsample}.  For each down-sampling channel, we list the number of filter groups, as well as each of their upper-bound frequency (i.e., $f_J$ for group $J$), and the computational cost as computed by Eq.~\eqref{CCrecom}.   Computational cost here is measured by MFLOPS, or $10^6$\,FLOPS.
    \label{tab_recomb}
}
  \end{center}
\end{table*}

Assuming our frequency division is made in a way such that each filter group has the same sample rate ($S_J$ for group $J$), then the  total recombination cost is 
\begin{equation}  \label{CCrecom:new}
  \mathcal{C}_{\mathrm{recom}}=\sum_J 8 S_J\,.
\end{equation}
In language of Sec.~\ref{subsec:implementation}, if we assume there are $N_{\mathrm{group},k}$ IIR filter groups for each down-sampling channel, then the recombination cost can also be written as 
\begin{equation}  \label{CCrecom}
  \mathcal{C}_{\mathrm{recom}}=\sum_k 8 S_k N_{\mathrm{group},k}\,.
\end{equation}
As a consequence, assuming that $\Delta A_{\rm cb} = R \Delta A_{\rm fb}$,  we have a total cost of
\begin{eqnarray}
\mathcal{C}_{\rm total} &=& 
\mathcal{N}_{\rm fb} \left[\frac{\mathcal{C}_{\rm IIR}}{R} +
\left(1-\frac{1}{R}\right)C_{\rm recom}\right]  \nonumber\\
&\approx& \mathcal{N}_{\rm fb} \sum_k \left[\frac{12 N_{\mathrm{IIR},k}}{R} 
+8N_{\mathrm{group},k}\right] S_k\,,
\end{eqnarray}
with the approximation valid when $R \gg 1$.  In this case, we can have a good estimate of the computational cost of IIR filtering with interpolation.  For a coarse bank with density $1/R$ the fine bank, filtering cost naturally decreases to $1/R$ of the cost of conventional IIR filtering without interpolation.   The cost of recombination can be estimated with a simple rule: for each sample rate, the  cost of recombination is about $2/(3\bar n_k)$ times that of conventional IIR filtering, where 
\begin{equation}
\bar n_k \equiv \frac{N_{\mathrm{IIR},k}}{N_{\mathrm{group},k}}
\end{equation}
is the average number of IIR filters in groups at the $k$-th sample rate.    As a consequence, the total cost of the IIR filtering with interpolation scheme including recombination can be lowered significantly {if we achieve a balance of $R \gg 1$ and $\bar n \gg 1$.  Note larger $R$ means larger coarse-bank grid size $\Delta A_{cb}$ for a fixed $\Delta A_{fb}$.  This is achieved by introducing finer frequency intervals.  On the other hand, finer frequency intervals means more IIR groups $N_{\mathrm{group}}$ or smaller $\bar n$ within each down-sampling channel. 

The computational cost for performing down- or up-sampling is implementation-dependent (see discussions in \cite{LLOID}).  They are not included in our calculation for simplicity.   We only need to perform data downsampling once for all templates, so the cost should be negligible compared to the total cost.   The upsampling process is needed at least for each coarse-bank template, but only for filter group outputs.  Note the number of filter groups is much smaller than the total number of the IIR filters.  Depending on the type of upsampling filters,  the upsampling cost  can be negligible compared to the total cost, but can also be in similar orders as the recombination cost.  This requires further investigation.

\subsection{Implementation for initial, Advanced LIGO and Einstein Telescope}


We first investigate the case of initial LIGO {to demonstrate the feasibility of our interpolation method}.  Taking into account the fact that even the optimal match between IIR filter and the real signal is not unity, { we need to place the fine-bank IIR template a little denser than that from theoretical waveform.  Theoretically for the Newtonian waveform,  we have $\Delta A_{\rm fb} = 923$ (in units of s$^{-5/3}$)  to have a minimum match of 0.97 for templates based on the signal waveform.  For the IIR filter bank, } we need a smaller spacing of $\Delta A_{\rm fb}^{\rm IIR} =800$ s$^{-5/3}$ in order for the bank to achieve the same match between an IIR template and the signal at $A+\Delta A_{\rm fb}^{\rm IIR}$.   { Fig.~\ref{ovp} shows numerically calculated match as a function of template spacing $\Delta A$ for templates from the signal waveform (black solid line) and for the IIR filters (red solid line) for the case of (1.4+1.4) \msun\  binary. }  Note that the numbers of fine-bank templates here are slightly different from those given in Table~\ref{basic_info}, as we use slightly different overlap and also we use numerically evaluated matches here, instead of ones computed analytically assuming high match (in Sec.~\ref{templatebank}).

To test the coarse-bank template placement, for simplicity, we restrict ourselves with the case of subdividing the frequency band into a total of six segments (or equivalently,  six IIR filter groups in the time domain).  According to the idealized theoretical calculations in frequency domain (Sec.~\ref{NCintp}), the optimal frequency subdivision {predicts} $\Delta A_{\rm cb}/\Delta A_{\rm fb} \approx  26$ for a minimum match of 0.97.  This calculation has assumed high match, and divides signals into parts that are strictly localized within separate frequency bands.  On the other hand, the numerical result using interpolation method on the IIR filter groups in the time domain (as prescribed in Sec.~\ref{IIRintp}) reveals that we can relax the coarse-bank spacing up to $\Delta A_{\rm cb}^{\rm IIR} = 19845$ s$^{-5/3}$  (dashed curve in Fig.~\ref{ovp}), meaning
\begin{equation}
\frac{\Delta A_{\rm cb}^{\rm IIR}}{\Delta A_{\rm fb}^{\rm IIR}} \approx 25.
\end{equation}
This is in very good agreement with the idealized prediction.  Fig.~\ref{ovp} shows in dashed line the numerical result of the match as function of $\Delta A$ for the interpolated IIR filtering method. 

We can now evaluate the total  computational cost of the entire filtering-reconstruction process.  For filtering, since we only have 
\begin{equation}
\mathcal{N}_{\rm c} =(A_{\rm max}-A_{\rm min})/(2 \Delta A_{\rm cb}^{\rm IIR}) = 92
\end{equation}
templates in the coarse bank~\footnote{Recall that since match between template at $A$ and signal at $A+\Delta A$ is already satisfactory, the next template needs to be placed at $A+2\Delta A$.}, and the cost for each full filtering is 2.4\,MFLOPS (see Table~\ref{tab_downsample}),  the cost of IIR filtering is $\mathcal{C}_{\rm IIR}^{\rm bank} = 221$\,MFLOPS. Since the number of templates in the fine bank is  
\begin{equation}
\mathcal{N}_{\rm f} =(A_{\rm max}-A_{\rm min})/(2 \Delta A_{\rm fb}^{\rm IIR}) = 2281\,,
\end{equation}
while the reconstruction cost for each member is 0.10\,MFLOPS, the total cost for reconstruction (for members in the fine bank but not already in the coarse bank) is 228\,MFLOP. Therefore the total cost for searching for Newtonian Chirps in initial LIGO is  449\,MFLOPS, or 0.5\,GFLOPS. 

We carry out the same procedure for Advanced LIGO and ET$_{\rm B}$, with frequency division information listed in Table~\ref{tab_recomb}, and interpolation factor as well as break-down of filtering and recombination costs listed in Table~\ref{tab_final}.  As we can read from Table~\ref{tab_final}, the computational power required for a real-time search of Newtonian Chirps, using IIR filters and interpolation,  in initial, Advanced LIGO and ET are 0.5\,GFLOPS, 1.2\,GFLOPS and 4.4\,GFLOPS, respectively. The scaling of cost with $f_{\rm min}$ is rather mild as expected, and the cost, even for ET, seems very manageable. 

In summary, it seems possible that to search for tens to hundreds of thousands of fine-bank templates for advanced LIGO or ET,  we can have the entire search done with a few desktop computers and fewer if other acceleration technique  such as the Graphics Processing Unit~\cite{shinkee10, shinkeeConf10} can be adopted. While our result is based on the Newtonian chirp, this outcome should be applicable to Post-Newtonian (PN) cases.  Note the low-latency pipeline LLOID with the FIR scheme in combination with downsampling and SVD technique \cite{LLOID, svd} also predicts manageable computing power for Advanced LIGO.   MBTA method \cite{MBTA}, on the other hand,  can already perform network analysis to search for inspiral signals using PN waveforms with a few CPUs for the initial LIGO.     How it scales with advanced detectors while maintaining low latency remains to be investigated (see also Sec.~\ref{sec:tf} for a comparison of frequency vs time domain method).   The integration of the time-domain IIR filtering method with the infrastructure of the LLOID pipeline is currently under way.  Preliminary result for the application of the IIR filterbank method to PN waveforms can be found in~\cite{hooperConf10} and ~\cite{hooper11}.

\begin{table*}
\begin{tabular}{|c||c|c||c|c||c|c||c|c||c|c|}
\hline
& $\Delta A_{\rm fb}^{\rm IIR}$ &  $\mathcal{N}_{\rm fb}$ &  $\Delta A_{\rm cb}^{\rm IIR}$ & $\mathcal{N}_{\rm cb}$ & $\mathcal{C}_{\rm IIR}$& $\mathcal{C}_{\rm IIR}^{\rm total}$ & $C_{\rm recomb}$ &$\mathcal{C}_{\rm recomb}^{\rm total}$ & $\mathcal{C}_{\rm total}$ & $\mathcal{C}_{\rm total}/\mathcal{C}_{\rm total}^{\rm fb}$\\
\hline
iLIGO & 800 & 2281 & 19845 & 92 & 2.4 & 221 & 0.10 & 228 & 449 & 0.082 \\
\hline
aLIGO & 255 & 7156 & 10500 & 174 & 3.0 & 522 & 0.090& 628  &1150 & 0.054 \\
\hline 
ET & 70 & 26069 & 2713 & 673 & 3.3 &  2221 & 0.083 & 2108 & 4329 & 0.050 \\
\hline
\end{tabular}
\caption{Break-down of total computational cost {in MFLOPS} in searching for Newtonian Chirps in initial LIGO, Advanced LIGO and ET, assuming interpolation {for inspirals of  1--3 \msun\  individual masses}. Here we list numbers of templates in both the fine ($\mathcal{N}_{\rm fb}$) and coarse banks ($\mathcal{N}_{\rm cb}$), the computational cost for each full IIR chain ($\mathcal{C}_{\rm IIR}$, taken from Table~\ref{tab_downsample}), as well as the recombination cost for each template ($\mathcal{C}_{\rm recomb}$, taken from Table~\ref{tab_recomb}).  We then give the total IIR filtering cost ($\mathcal{C}_{\rm IIR}^{\rm total}$), the total recombination cost ($\mathcal{C}_{\rm recomb}^{\rm total}$), and the grand total cost. We also list the ratio $\mathcal{C}_{\rm total}/\mathcal{C}_{\rm total}^{\rm fb}$, in which $\mathcal{C}_{\rm total}^{\rm fb}$ represents computational cost for the full bank without using interpolation.
\label{tab_final}
}
\end{table*}

\section{Time domain vs frequency domain approach} 
\label{sec:tf}

{\subsection{General consideration}}
{In terms of template interpolation, the ideas to divide the template into segments in the time or frequency domain  are equivalent in mathematics -- both trying to represent the template by the superposition of a complete basis of the continuous real-value function space on real axis. 
The functions in the basis are much simpler than the template, and thus easier to deal with. We can improve the computational efficiency by processing the basis functions first and then superpose them in the right way to get the result for a template. Given that the data we get from the detector is in the time domain, the advantage of working in the time domain is that we can avoid procedures required to transform the data into frequency domain (e.g., data accumulation in Fourier transformation) and easily achieve low time latencies.  On the other hand, working in frequency domain allows us to easily combine the algorithm with down sampling technique and reduce the number of templates.}

The frequency-domain template interpolation technique, e.g., that used in MBTA\cite{MBTA},  usually uses Heaviside function to cut the template. So the template can be superposed smoothly in the  frequency domain while in the  time  domain the joint of different basis functions can be quite crude.  This means that those methods with this technique could easily take advantages of working in the frequency domain, but not  both in the time and frequency domain without substantial additional cost in computation. 

{
Our algorithm, with IIR filters working in the time domain and template interpolation designed from the frequency domain, takes advantages of the benefits from both the time and frequency domain approach.  Because we use a relatively smooth cut in both domains, we can both achieve low latency in the time domain and reduce the total number of templates while taking advantages of the down sampling technique.
}

 {\subsection{Comparison of computational efficiency \label{Fourier Transform}}}

When latencies of the analysis are not in concern, the frequency domain implementation of the cross correlation of data with templates (Eq.~\ref{matchedFilter}) is probably the most computationally efficient approach.   This is due to the use of Fast Fourier Transform technique that has $O(N\log N)$ operation count ($N$ is the number of data points) as compared to the $O (N^2)$ operation count for the FIR method described previously.  On the other hand, the operation count of the IIR filterbank method is $O(N)$ but multiplied with a coefficient directly related to the possibly large number of filters needed to achieve a desired match to the chirp signal.   Here we  take latencies into consideration and compare the computational efficiency of the FFT-based method with the proposed IIR method.




To obtain low-latencies for the FFT-based matched filtering prescribed in Eq.~\eqref{matchedFilter},  the most straightforward approach is to analyze data in overlapping segments. We consider the analysis of equal-length segments  of duration $T_{\rm stretch}$ as shown in Fig~\ref{Fourier_stretch} with the duration of overlap equal to that of the longest signal, and the rest termed
$T_{\mathrm{latency}}$, that is,
\begin{equation}\label{Tstretch}
  T_{\rm stretch}  = T_{\rm longest} + T_{\rm latency} .  
\end{equation}
Here we assume the same strategy as in the current GW search pipeline where FFTs are performed with fixed length that accommodates the longest signal to ensure the coverage of signals of all possible duration.  Note in practice, longer $T_{\rm stretch}$ might be needed to take into account of the windowing effect of the FFTs and issues like the sharp notch filter problems due to lines in the noise power spectrum \cite{bruce05}.  For each data stretch, the output of Eq.~\eqref{matchedFilter} has also the duration $T_{\rm stretch}$, but due to the wrap-around effect of FFTs, only outputs (for signals with ending time) within the last  $T_{\rm latency}$ are valid. This means that to obtain a valid output of duration $T_{\rm latency}$,  a data stretch of at least $T_{\rm longest} + T_{\rm latency}$ needs to be processed. 
\begin{figure}
  \includegraphics[width=3in]{{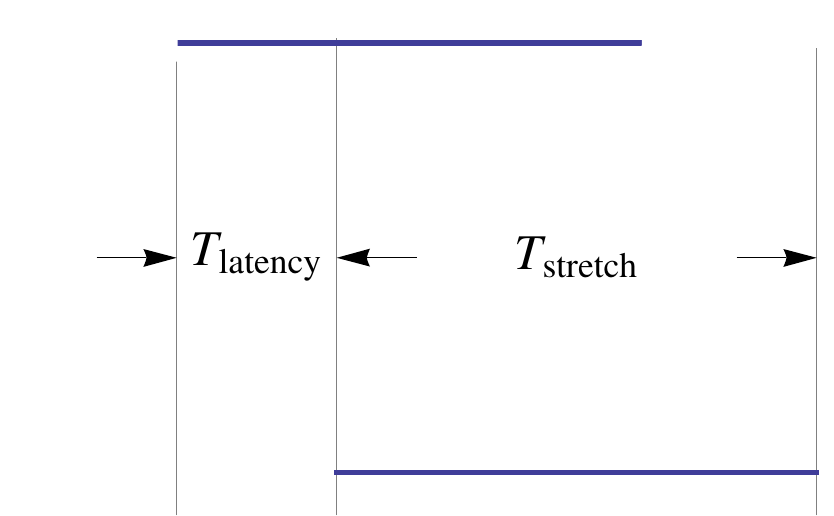}}
  \caption{\label{Fourier_stretch} Analysis with overlapping data
    segments.   The two horizontal lines
    represent two adjacent data stretches used for FFT.  The lower data segment starts data accumulation with a delay of $T_{\rm latency}$ relative to the upper one.  The duration of the overlap
    between the two stretches is that of the longest signal in the template
    bank (see text in Sec.~\ref{Fourier Transform}).}
\end{figure}
%
The requirement to perform filtering in real-time implies that the entire analysis needs to be completed within $T_{\rm latency}$ seconds. The minimum total number of real multiplications and real additions for the FFT algorithm is about $4N\log_2N$ \cite{johnson07, lundy07}.  Therefore the minimum computational cost in terms of FLOPS for each template for a real-time FFT-based matched filtering is at least,
 
\begin{equation}\label{FourierCC}
  \mathrm{C}_{\rm FFT} = \frac{4 S\cdot T_{\mathrm{stretch}}\log_2 (S\cdot T_{\mathrm{stretch}})} {T_{\mathrm{latency}}}, 
\end{equation}
where $S$ is the data sampling rate. Here we assume a uniform sampling rate.


\begin{figure}[htb]
  \centering
  \includegraphics[width=\columnwidth]{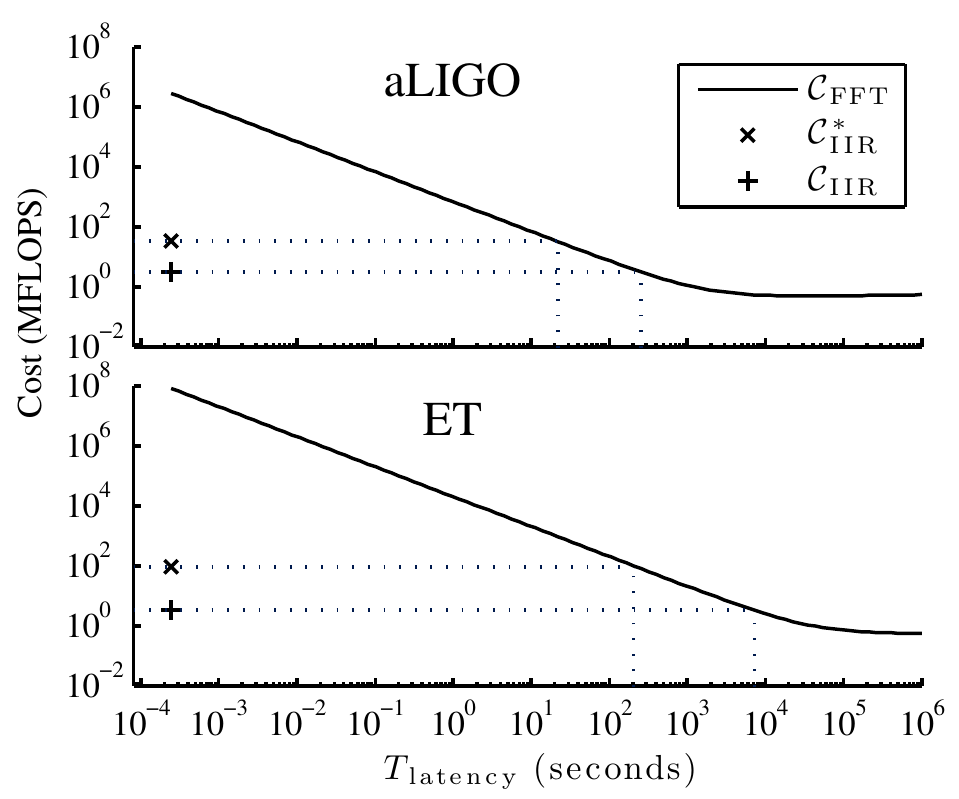}
  \caption{Computational cost as a function of $T_{\rm latency}$ for a straightforward FFT analysis with overlapping data segments (solid line) and for the IIR filter method with downsampling technique (``+" symbols) and without (``$\times$" symbols ) for real-time filtering with one template of a (1.4+1.4) \msun\  binary.    The upper panel shows the cost for aLIGO and the lower one for ET.  The dotted lines illustrate the equal cost between the FFT and IIR method and the corresponding latencies. 
The computational cost of the FFT method is calculated from Eq.~\eqref{FourierCC} with  the longest template taken to be that of (1+1) \msun\  binary and sampling rate $S$=4096Hz. The IIR data are from Table~\ref{tab_downsample} (column 15) (with downsampling)  and Eq.~\eqref{IIRCC} (without downsampling).   }
\label{figFLOPS}
\end{figure}
In the FFT method, the actual delay $T_{\mathrm{delay}}$ between the end time of a GW signal and
the event triggering depends on where the signal lies
in the data stretch. The longest delay occurs when the ending time of
a signal lies $(T_{\rm latency} - dt)$ before the end of a data
stretch where $dt\equiv 1/S$ is the sampling interval. In this case, after the signal ends, it takes the segment further $(T_{\rm latency} - dt)$ time to finish
accumulating data, and then another $T_{\rm latency}$ to be processed, resulting in a
delay of,
\begin{equation}
 T^{\mathrm{FFT},\mathrm{worst}}_{\mathrm{delay}} = 2T_{\mathrm{latency}}-dt \approx 2T_{\rm latency}  
\label{FFT_worst}
\end{equation}

The shortest latency is achieved when the ending time of a signal lies just at the end
of the data stretch, in which case the waiting time for the data to be
analyzed is zero and the delay time of obtaining the trigger is simply the
analysis time,
\begin{equation}
 T^{\mathrm{FFT},\mathrm{best}}_{\mathrm{delay}} = T_{\mathrm{latency}}.  
\label{FFT_best}
\end{equation}
Therefore, for the FFT method, the delay time between the end of the signal and the event triggering is about 1--2 times of $T_{\mathrm{latency}}$.  Although in previous LIGO inspiral search pipelines $T_{\rm latency}$ is usually chosen
so that adjacent data stretches are overlapped by 50\%, it can be chosen so
that $T_{\rm latency}$ is much smaller, meaning data segments  are analyzed with larger overlaps and higher computational cost.  

In comparison, for the IIR method, every new data point will be processed
immediately when it is available. The delay time between the end of the signal
and the triggering time can therefore in principle be as small as the data
sampling interval.  For real-time processing, the analysis time of the IIR filters at each time step should also be within one sampling interval, $dt$.  As discussed previously, for each output of an IIR filter in \eqref{iirfilter}, a total of 12 floating point operations are needed.  Hence to produce the IIR filter bank output in real time without downsampling  for one template requires the floating point operation per unit time of  
\begin{equation}
  \label{IIRCC}
  \mathrm{C}_{\rm IIR}^{*} = 12 S \cdot N_{\rm IIR},
\end{equation}
and the delay 
\begin{equation}
 T^{\mathrm{IIR}}_{\mathrm{delay}} = dt.
\label{IIR_delay}
\end{equation}
Here asterisk is used to indicate the computational cost of the IIR filter method without the downsampling technique. 

Fig.~\eqref{figFLOPS} shows the computational cost of one template for the FFT method as a function of $T_{\rm latency}$ when searching for a GW from a (1.4+1.4) \msun\  NS-NS binary and its comparison to that of the IIR filter method with and without downsampling technique.  It shows that the computational cost of the FFT based method increases as latencies decreases,  the increase is particularly significant at latencies less than hundreds to thousands of seconds  (Eq~\eqref{FourierCC}), whereas IIR methods (Eq.\eqref{CCiir}, Eq.~\eqref{IIRCC}) have an inherent latency of the sampling interval (i.e \textbf{not} a function of latency).  It is clear that the IIR filter method presented in this paper has significant advantage over the FFT method in computational efficiency when low latencies  are in demand.  In particular, for Advanced LIGO,   the IIR method can be much more efficient at latencies less than  a few $\times 10^2$ seconds.   For the Einstein Telescope, IIR filter method can be much more  efficient at latencies less than a few $\times 10^3$ seconds.

It should be mentioned that we compare only the core computational cost for the IIR and the FFT method for one template.   We purposely leave out the cost of whitening or the cost to take care of other FFT effect such as windowing effect as they are very much implementation-dependent.   We also do not include template interpolation method for both methods as they are very much implementation-dependent.   In practice, both methods require that the raw data be conditioned, transported, pre-whitened before they are ready to be analyzed.  These are expected to cause additional latencies on the order of tens of seconds.

\section{Conclusion\label{conclusion}}

In this paper, we show that a time-domain search algorithm, with the flexibility of being able to detect a (non-precessing) compact-binary coalescence even before the final merger, is not only feasible for advanced and even future ground-based gravitational-wave detectors --- but in fact can be realizable by a small number of state-of-the-art personal computers.  

In addition to employing the multi-rate technique for time-domain filtering, we have developed two additional key techniques in order to bring down the computational cost into the realm of feasibility:  (i) the conversion of a chirp signal into a chain of IIR filters, and (ii) an algorithm that allows the reconstruction of filtering results of a finely spaced template bank from a much coarser bank, when each template in the coarse bank is divided into sub-templates.  In order to illustrate the main techniques, we have restricted ourselves to the Newtonian Chirp, but it is rather straightforward to generalize our algorithms into post-Newtonian templates. 

Our main results on computational cost of the time-domain algorithm, for initial, advanced and future detectors, are summarized in Table~\ref{tab_final}.  With a simple comparison, we also conclude that our time-domain algorithm should require less computational resources than the conventional frequency-domain approach, when a short latency of less than hundreds to thousands of seconds is required --- as shown in Fig.~\ref{figFLOPS}.

Besides being computationally efficient at low (or even negative) latencies, the IIR filter bank method is also much simpler to implement than the FFT-based methods, making it ideal for parallel computing, e.g., with Graphics Processing Units~\cite{shinkee10}.

Two further ingredients must be added into the search pipeline before we can set up an early-warning system for EM follow-ups of compact binary coalescence: (1)  a reliable veto strategy, and (2) an efficient algorithm for sky localization. The fact that our numerical results for IIR filter groups agree so well with frequency-domain analytical estimates (Sec.~\ref{subtemplate}) assuming sharp divisions in frequency indicates that the sub-IIR-groups can be well-approximated as independent contributions to the SNR.  This means a $\chi^2$-like test that compares relative SNR contributions from filter subgroups to their expectations would be a promising veto strategy (see also \cite{cannon11} for other strategies that might be applicable for further efficiency improvement.)

As for localization, we could in principle adopt the existing algorithm already in place in the LIGO/VIRGO pipeline, which is based on coincidences of SNRs among multiple detectors.  Alternatively, the fact that IIR filters are separated in both time and frequency may provide a possibility of developing a {\it coherent} search pipeline with feasible computational cost. The reason for the high number of templates in a coherent search is directly due to the multiplication of the high number of templates 
along the direction of mass parameters and the high number of sky locations.  However, as we divide each template into frequency segments, we find that in low frequencies, although there is a large number of cycles, and hence a requirement for a finer separation in mass parameters, the sky resolution of a detector network is low and there does not need a high number of sky patches; in high frequencies, we need a fine grid in the sky, but a coarse grid in mass parameters.  As a consequence, we may need a much lower number of sub-templates are required for each frequency segment.  This is currently being investigated.  
 
\acknowledgments
{We are grateful for inspiring discussions with Rana Adhikari, Kipp Cannon, Chad Hanna, Drew Keppel, Alan Weinstein, Patrick Brady, Shin Kee Chung, David Blair,  Benoit Mours, Leo Singer, Peter Shawhan, and  Nickolas Fotopoulos.  }  This work has been supported in part by NSF Grants PHY-0601459, PHY-0653653, (LIGO) and CAREER Grant PHY-0956189 and the David and Barbara Groce start-up fund at Caltech, {and by ARC Discovery Project and ARC Future Fellowship program.}

\bibliographystyle{apsrev} 
\bibliography{references}



\end{document}